\documentclass[]{imag-ms-template}

\usepackage{array}
\usepackage{subcaption}
\usepackage[switch]{lineno}
\usepackage{booktabs}
\usepackage{multirow}
\usepackage{pdfpages}
\usepackage{xcolor}
\usepackage{soul}
\sethlcolor{yellow}

\newif\ifshowrev
\showrevfalse     % highlights OFF

\newcommand{\rev}[1]{%
  \ifshowrev
    \hl{#1}%
  \else
    #1%
  \fi}
\title{Midsagittal Corpus Callosum Parameterization: A Framework for Spatially Detailed Diffusion MR Imaging Analysis}

\author{Caio Santana,$^{1,2,4,5,6,\ast}$ Catherine Lebel,$^{2,3,4}$ Richard Frayne,$^{2,4,5,6}$ Letícia Rittner$^{1}$\\
{\small $^{1}$School of Electrical and Computer Engineering, Universidade Estadual de Campinas (UNICAMP)},\\
{\small Campinas, SP, Brazil}\\
{\small $^{2}$Department of Radiology, University of Calgary, Calgary, AB, Canada}\\
{\small $^{3}$Alberta Children's Hospital Research Institute, University of Calgary, Calgary, AB, Canada}\\
{\small $^{4}$Hotchkiss Brain Institute, University of Calgary, Calgary, AB, Canada}\\
{\small $^{5}$Department of Clinical Neurosciences, University of Calgary, Calgary, AB, Canada}\\
{\small $^{6}$Seaman Family MR Research Centre, Foothills Medical Centre, Calgary, AB, Canada}\\
{\small $^\ast$Correspondence: c218653@dac.unicamp.br}\\
{\small Data used in preparation of this article were obtained from the Alzheimer's Disease Neuroimaging}\\
{\small Initiative (ADNI) database (adni.loni.usc.edu). As such, the investigators within the ADNI}\\
{\small contributed to the design and implementation of ADNI and/or provided data but did not participate}\\
{\small in the analysis or writing of this report. A complete listing of ADNI investigators can be found at:}\\
{\small http://adni.loni.usc.edu/wp-content/uploads/how\_to\_apply/ADNI\_Acknowledgement\_List.pdf}
}

\begin{document} 

% \linenumbers

\maketitle 


\begin{abstract}

The corpus callosum (CC) is the largest white matter structure in the human brain and plays a central role in interhemispheric communication. Diffusion MR imaging is widely used to investigate microstructural organization, but commonly used analysis approaches have important limitations. Voxel-based methods rely heavily on accurate inter-subject registration and smoothing choices, tract-based spatial statistics (TBSS) restricts the analysis to a narrow white matter skeleton, and parcellation approaches reduce spatial detail by averaging diffusion measures over relatively large regions.
In this work, we introduce a parameterization framework for the analysis of the midsagittal CC that resamples the structure using a standardized grid of points, \rev{establishing a geometric correspondence across individuals} without explicit inter-subject registration.
In a test-retest dataset, the method provided stable measurements while covering most of the midsagittal CC, \rev{and sensitivity analyses showed that results in the central CC changed little under perturbations of the input CC mask, whereas peripheral regions were more sensitive.}
The framework was further evaluated in group analyses of aging and Alzheimer’s disease \rev{in datasets that differ in acquisition protocol and spatial resolution. Effect-size maps and quantitative comparisons showed that its results were largely consistent with TBSS and parcellation-based analyses in the overall spatial pattern of effects, while also revealing additional spatially localized differences.}
These findings suggest that the parameterization provides a simple, computationally efficient, and spatially detailed approach for investigating midsagittal CC microstructure, \rev{offering a structure-specific complement to whole-brain and region-based methods.}

\end{abstract}

% ===========================================================
\section{Introduction}
\label{sec:introduction}
% ===========================================================

The corpus callosum (CC) is the largest white matter (WM) structure in the human brain and the primary communication pathway between the cerebral hemispheres~\citep{aboitiz_one_2003}. Composed of highly packed and organized neural fibers (axons), the CC is located in the middle of the brain of placental mammals and extends from the midline toward the cerebral hemispheres through its radiations. These radiations connect many regions of both hemispheres, facilitating the integration and coordination of cognitive, sensory, and motor signals~\citep{shah_corpus_2021, turgut_corpus_2023}.

The development of the CC begins around 8 weeks of gestation, and its characteristic shape can be observed from 18 to 20 weeks~\citep{rakic_development_1968}. While all callosal fibers are typically present at birth, the CC continues to undergo structural changes due to myelination, fiber redirection, and pruning. In particular, its postnatal growth is attributed to the myelination process, which follows a posterior-to-anterior direction, with the myelination of posterior primary cortical areas preceding that of the anterior associative regions~\citep{turgut_corpus_2023}. The CC continues to grow in size until reaching a plateau in the third decade of life and begins to show age-related volumetric decline around 60 years of age~\citep{danielsen_lifespan_2020}.

The morphological and microstructural changes the CC undergoes throughout life have a critical impact on brain function~\citep{rakic_development_1968, turgut_corpus_2023, danielsen_lifespan_2020}. Therefore, understanding its development and subsequent age-related decline is of high clinical relevance. Sex-related differences in the CC have also been reported in multiple studies~\citep{witelson_hand_1989, tanaka-arakawa_developmental_2015, turgut_corpus_2023}. Moreover, given its role in interhemispheric communication, the CC is a common target for study over a wide range of neurological and mental health conditions including obsessive-compulsive disorder~\citep{piras_corpus_2021,haghshomar_white_2023}, schizophrenia~\citep{piras_corpus_2021,arnone_meta-analysis_2008}, multiple sclerosis~\citep{caverzasi_mwf_2023,tabibian_corpus_2025}, and Alzheimer’s disease (AD)~\citep{sexton_meta-analysis_2011,kumar_topology_2022}.

% -----------------------------------------------------------
\begin{figure}[b!]
\centering
\includegraphics[width=0.5\linewidth]{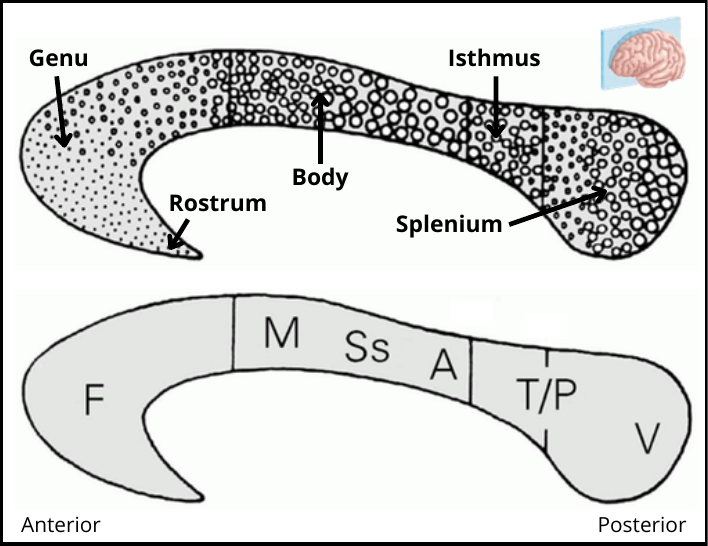}
\caption{Representation of anatomical parts (arrows) and fiber diameters along the midsagittal section of the CC (top). Subregions of the CC associated with different cortical brain areas (bottom): F, frontal; M, motor; Ss, somatosensory; A, auditory; T/P, temporoparietal; V, visual. Adapted from~\citet{aboitiz_one_2003}.}
\label{fig:CC_subdivision}
\end{figure}
% -----------------------------------------------------------

Anatomically, the CC is commonly subdivided into five principal segments: the rostrum, genu, body, isthmus, and splenium (Fig.~\ref{fig:CC_subdivision}). These subdivisions exhibit distinct developmental trajectories and have unique microstructural characteristics~\citep{rakic_development_1968}. Histological analyses have demonstrated regional variations in axon diameter, fiber density, and degree of myelination across CC segments~\citep{aboitiz_fiber_1992}. Callosal fibers are topographically organized according to the cortical areas they connect, reflecting functional specialization. The more anterior regions primarily mediate the transfer of motor and higher-order perceptual information between frontal brain regions, whereas more posterior segments are involved in the integration of visual, auditory, and somatosensory information~\citep{aboitiz_one_2003,fabri_functional_2013}. This organization is clearly depicted in the midsagittal plane (midline), where callosal fibers are represented in a compact two-dimensional form. This simplified representation reduces anatomical and analytical complexity, enabling reproducible measurements and facilitating comparisons across individuals and populations. For this reason, the midsagittal section of the CC is extensively studied. Notably, studies have indicated an association between the midsagittal area of the CC and the total number of callosal fibers, supporting the biological relevance of this representation~\citep{aboitiz_fiber_1992,riise_stereological_2011}.

Several imaging techniques can be employed to study the CC \textit{in vivo}. Conventional structural magnetic resonance (MR) imaging is widely used to characterize callosal morphology, particularly T1- and T2-weighted (T1w and T2w) sequences. Global morphological analyses assess CC size, shape, curvature, and overall geometry, providing insights into developmental trajectories and structural alterations associated with neurological and psychiatric conditions~\citep{garel_biometry_2011,turgut_corpus_2023}. Thickness measurements quantify regional variations along the CC~\citep{piras_corpus_2021,adamson_thickness_2011}, providing a more localized description than global metrics alone. Additionally, owing to its size and characteristic midsagittal organization, the CC is often divided (or parcellated) into subdivisions to extract region-specific measures~\citep{witelson_hand_1989,hofer_topography_2006}.

Beyond structural characteristics, there is much interest in assessing the microstructural organization of the CC, as these measures can reflect the integrity and maturation of callosal fibers. By measuring the movement of water molecules within tissue, diffusion MR imaging (dMRI) provides indirect information about WM microstructure, enabling detailed assessment of the CC~\citep{hofer_topography_2006}. Diffusion tensor imaging (DTI) is the most widely used model for fitting dMRI data, characterizing tensors that describe diffusion in three dimensions~\citep{basser_mr_1994}. Although the DTI model has well-known limitations, particularly in regions with crossing or complex fiber configurations~\citep{tournier_diffusion_2019}, it is suitable for studying the CC due to its highly organized fibers. Using DTI-derived scalar maps, local properties can be assessed and compared between individuals or groups. The most commonly used DTI scalar measures are fractional anisotropy (FA), mean diffusivity (MD), radial diffusivity (RD), and axial diffusivity (AxD).

\rev{DTI scalar measures are sensitive to several tissue properties simultaneously, such as axonal density, diameter, and myelination, complicating biological interpretation. Advanced diffusion models and acquisitions have been proposed to provide metrics that are more specific to tissue microstructure. Multi-compartment models such as neurite orientation dispersion and density imaging (NODDI)~}\citep{zhang_noddi_2012}\rev{ estimate the microstructural complexity of dendrites and axons, and fixel-based analysis~}\citep{raffelt_investigating_2017}\rev{ characterizes the density and cross-section of individual fiber populations within a voxel. Approaches such as AxCaliber~}\citep{assaf_axcaliber_2008}\rev{ and ActiveAx~}\citep{alexander_orientationally_2010}\rev{ aim to estimate axon diameter distributions and densities, and time-dependent diffusion techniques~}\citep{wang_time-dependent_2026}\rev{ enable the mapping of tissue microstructure at different spatial scales by varying the diffusion time. The CC has been a frequent target of these methods, as its coherently oriented fibers and regional variations in axonal properties provide a reference for comparing microstructural estimates. Diffusion-based estimates have reproduced regional patterns consistent with histology across the genu, body, and splenium~}\citep{alexander_orientationally_2010,barazany_vivo_2009,lee_time-dependent_2020}\rev{, including regionally specific age-related differences in axon diameter and density, which are most pronounced in the genu~}\citep{fan_age-related_2019}.

To enable investigation of the CC, a wide range of computational methods have been developed in the literature. Basic morphometric measures, such as area and volume, can be directly computed from the segmented CC. More advanced descriptors, however, require additional modeling. For instance, the length of the CC can be defined as the length of its centerline, a curve defined as connecting the anterior tip of the genu to the posterior end of the splenium and equidistant from the superior and inferior midsagittal callosal boundaries~\citep{park_relationships_2011}. Thickness can be estimated as the distance from this midsagittal centerline to the CC boundaries~\citep{luders_parasagittal_2006}. Alternatively, a scalar potential field can be computed within the midsagittal CC by solving Laplace’s equation with fixed values imposed on the superior and inferior boundaries, from which smooth trajectories (streamlines) connecting the two boundaries are derived. Thickness at a given location is then defined as the length of the streamline passing through that point~\citep{adamson_thickness_2011}. Other approaches for shape analysis span from landmark-based techniques~\citep{bookstein_midline_2002} to statistical shape modeling frameworks~\citep{joshi_statistical_2013}.

There are also a variety of strategies for the microstructural characterization of the CC using diffusion-derived scalar maps. These approaches can be broadly categorized into three types: voxel-based, region-based, and tract-based analyses. The main voxel-based approaches, including voxel-based morphometry (VBM), involve aligning (spatially registering) images to a template, segmenting them into tissue types, smoothing the segmented data, and applying statistical tests to identify voxelwise differences between groups~\citep{ashburner_voxel-based_2000}. Although VBM was originally designed to compare gray matter density, VBM-style analyses can also be applied to diffusion data, usually the FA map. This type of analysis is relatively easy to apply, investigates the whole brain, and does not require the prior definition of regions-of-interest. However, especially when using diffusion data, problems can arise due to misalignment. It is also not trivial to define the optimal degree of smoothing, a choice that can significantly affect the results obtained~\citep{van_hecke_dti_2016}.

Tract-based spatial statistics (TBSS)~\citep{smith_tract-based_2006} was developed to overcome the problems of standard registration algorithms in voxelwise analyses of FA images and is widely used in the analysis of diffusion data. TBSS steps involve: nonlinear registration to align FA images from all individuals, computing the mean image and applying thinning to create a mean skeletonized FA image, projecting the local FA maxima from each individual onto the skeleton, and then performing voxelwise statistical analyses in skeleton space. Although it reduces registration problems, the skeletonization approach cannot compensate for large misregistrations and analyzes only a relatively small portion of the WM, \rev{leaving most of the midsagittal CC outside the analysis}~\citep{van_hecke_dti_2016,bach_methodological_2014}.

Region-based approaches divide the CC and calculate the mean values of diffusion maps within each subdivision for analysis and group comparison. The geometric methods proposed by \citet{witelson_hand_1989} and \citet{hofer_topography_2006} are the most well-known and widely used in the literature. Based on the population averages from their respective studies, both proposed subdividing the CC into vertical sections using fixed proportions along the anterior-posterior length, such that each section connects specific cortical areas. However, because these divisions do not account for individual variability, the geometric approaches tend to be inconsistent across participants and may not accurately reflect true interhemispheric connectivity patterns~\citep{santana_corpus_2025}. In contrast, \citet{cover_data-driven_2017} and \citet{santana_dti-based_2023} proposed automated CC parcellation methods based on DTI data, grouping voxels with similar FA values and diffusion tensors, respectively. Other studies have used advanced tractography algorithms to subdivide the CC. For example, \citet{friedrich_relationship_2020} considered eleven cortical regions and applied tractography to identify eleven CC regions. While they provide a simplified and CC-specific framework, there is no consensus regarding CC parcellation schemes~\citep{santana_corpus_2025}, largely due to the absence of visible boundaries or clear anatomical landmarks delimiting its subregions. Moreover, parcellation-based analyses tend to lack fine detail, as they are restricted to mean values computed over relatively large regions.

There is currently no simple framework that enables spatially detailed diffusion analysis across the midsagittal CC while avoiding both heavy reliance on nonlinear registration and the loss of spatial resolution inherent to region-based approaches. In this paper, we propose a parameterization-based framework for the analysis of the midsagittal CC from dMRI data. The method builds upon established ideas from morphometric CC analysis to resample the structure using a standardized grid of sampling points, enabling direct point-by-point comparisons across individuals \rev{at corresponding geometric locations}. The approach is computationally efficient, avoids explicit nonlinear inter-subject registration and spatial smoothing, and provides a spatially detailed characterization of diffusion properties across most of the midsagittal CC.
\rev{We evaluated the framework in three publicly available datasets that differ in population, acquisition protocol, and spatial resolution.} In the next section, we describe the method and the experiments conducted to define its parameters, assess its repeatability, \rev{and evaluate its sensitivity to the input CC segmentation}. We then present the results of these experiments and the final method configuration, followed by two DTI-based group analyses (aging and Alzheimer’s disease), \rev{in which the parameterization is quantitatively compared with TBSS, VBM-style, and parcellation-based approaches.}

% ===========================================================
\section{Materials and Methods}
\label{sec:methods}
% ===========================================================

\subsection{Corpus Callosum Parameterization Method}

\rev{The CC segmentation masks are the starting point for the proposed framework. Any segmentation strategy can be used, as long as it provides accurate masks of the midsagittal section of the CC in dMRI space.} Given these masks, the parameterization method defines a set of points along the CC to allow resampling into a standardized space, \rev{so that every individual is represented on the same grid of points regardless of variations in the shape and size of the CC, establishing a geometric correspondence between them.} Assuming these sampling points reflect comparable anatomical locations across individuals, the parameterization enables point-by-point comparison between individuals and groups.

\subsubsection{Boundaries Definition}

The method first identifies the midsagittal CC landmarks, namely its external boundary and the anterior (genu) and posterior (splenium) endpoints. From a binary mask of the CC midsagittal section, an initial estimate of the external boundary is obtained as an ordered list of contour voxels. A B-spline is then fitted to this contour, with a smoothing factor proportional to the number of boundary voxels to ensure a consistent level of regularization across individuals. Based on empirical evaluation, a smoothing factor of one-tenth of the number of boundary voxels was adopted, yielding a sufficiently smooth contour while preserving the reference point at the genu.

\begin{figure}[tb!]
  \centering
  \begin{tabular}{@{} m{0.8cm} @{} m{7cm} @{}}
    \small \textbf{(a)} & \includegraphics[width=7cm]{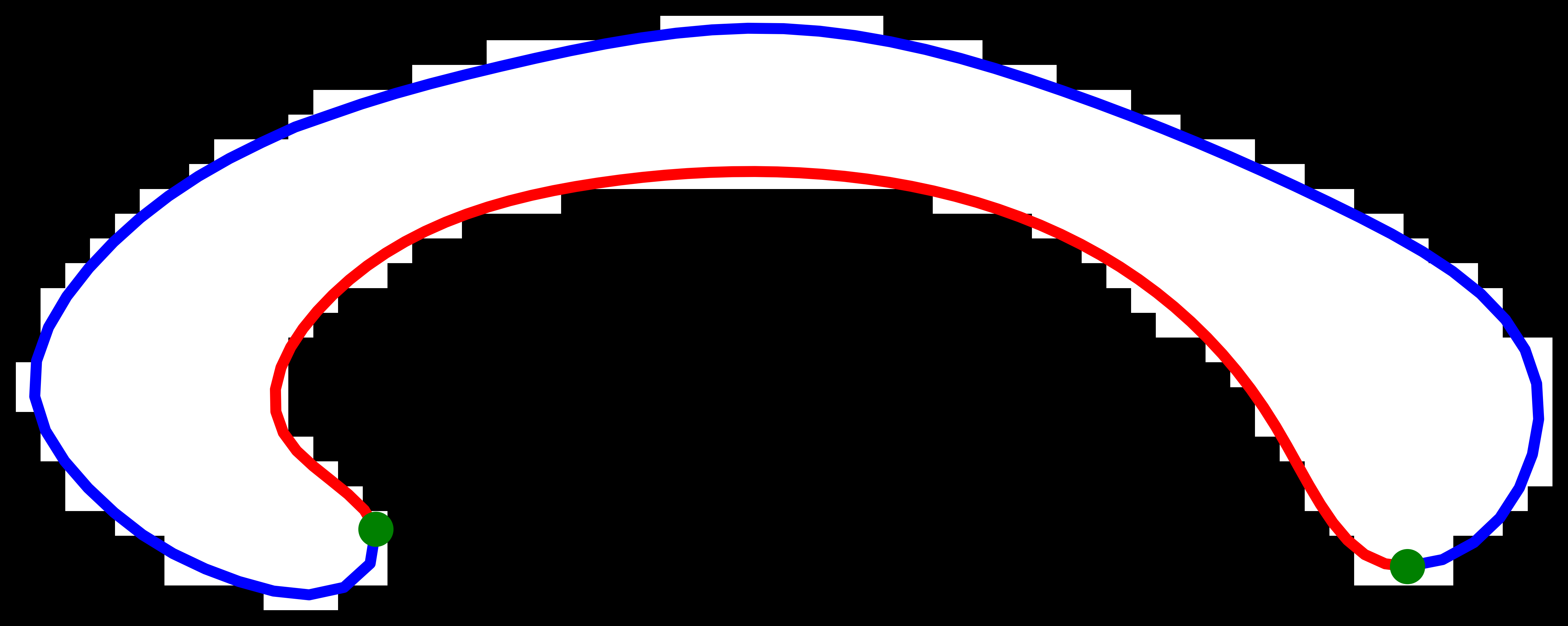} \\
    \small \textbf{(b)} & \includegraphics[width=7cm]{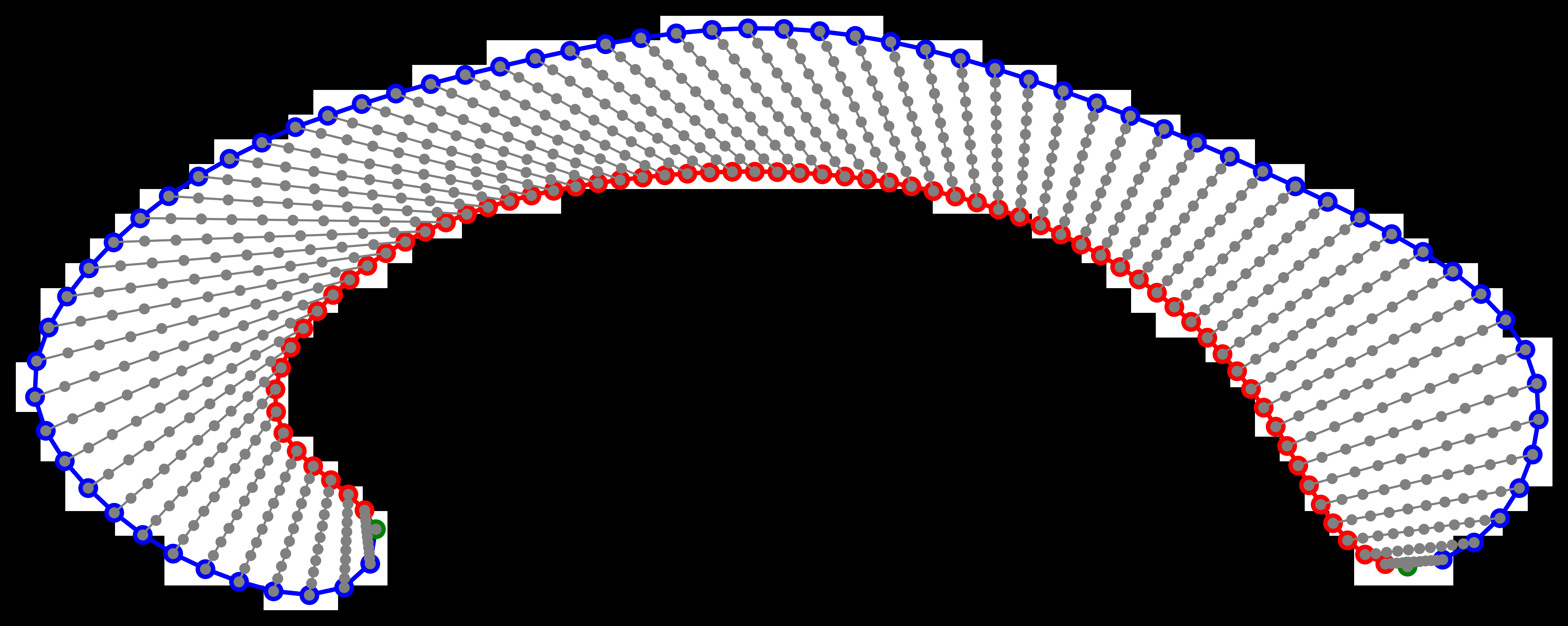} \\
    \small \textbf{(c)} & \includegraphics[width=7cm]{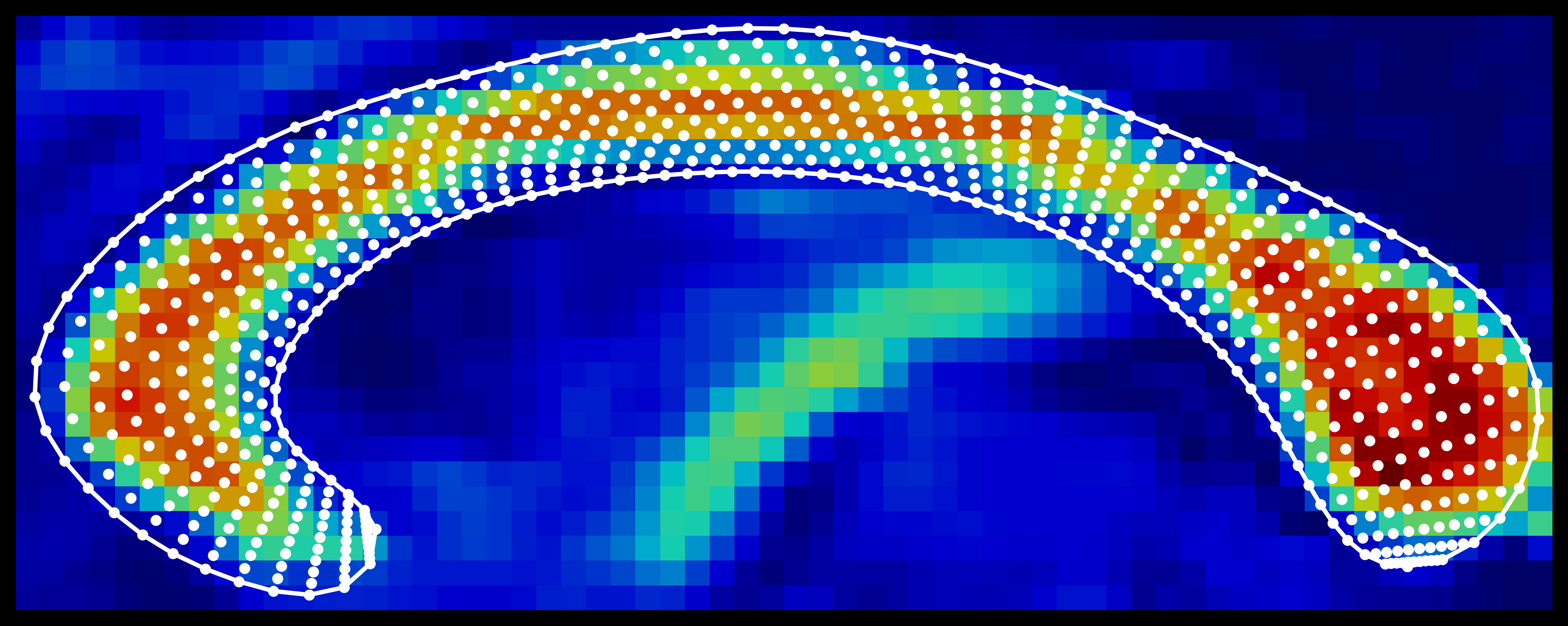} \\
    \small \textbf{(d)} & \includegraphics[width=7cm]{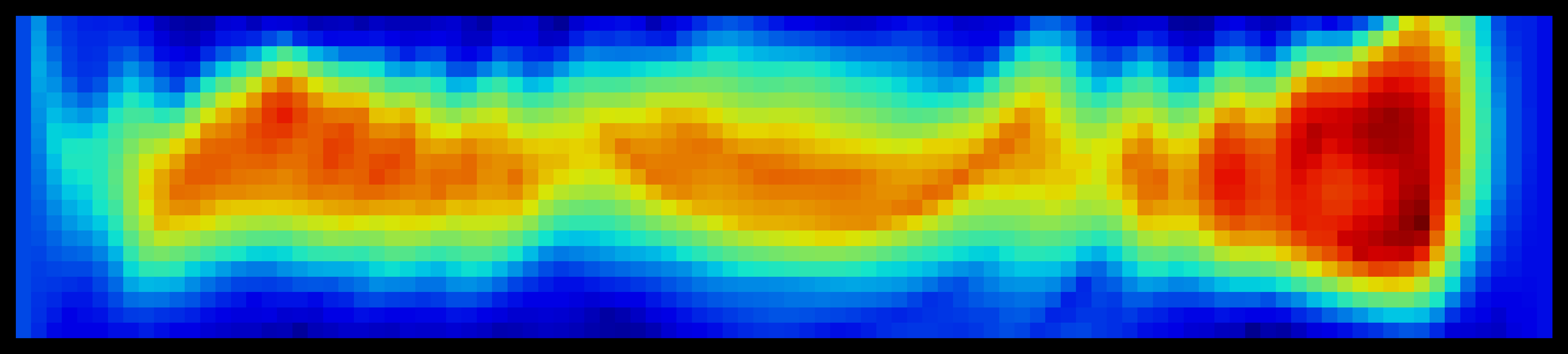}
  \end{tabular}
\caption{Corpus callosum parameterization method. \textbf{(a)} Superior (blue) and inferior (red) boundaries defined by the genu (left) and splenium (right) reference points (green).  Sampling points (straight lines method, see text) over \textbf{(b)} corpus callosum mask and \textbf{(c)} FA map. \textbf{(d)} Parameterized FA map in standardized image space.}
\label{fig:CC_param}
\end{figure}

To obtain a more refined representation, the fit spline was sampled at five times the number of boundary measurements. The resulting smoothed and upsampled contour was then divided into superior and inferior boundaries using the genu and splenium reference points (Fig.~\ref{fig:CC_param}a). Due to the characteristic ``C''-shape of the CC, these endpoints can be automatically identified~\citep{park_relationships_2011}:

\begin{itemize}

    \item \textbf{Anterior endpoint:} using the first and second derivatives of the spline, the curvature is computed within a restricted anterior region, and the point of maximum curvature is identified as the tip of the genu.
    
    \item \textbf{Posterior endpoint:} the reference point in the splenium is defined as the most inferior point in the posterior half of the CC.
    
\end{itemize}

Finally, separate splines were fit to the superior and inferior boundaries without smoothing.

% -------------------------------------------------
% -------------------------------------------------

\subsubsection{Internal Parameterization}

To map diffusion properties within the CC to standardized image space, an internal coordinate system was established using $N_b$ transverse curves connecting the upper and lower boundaries. We tested two different approaches to define these curves:

\begin{itemize}

    \item \textbf{Straight lines:} each boundary spline is resampled using $N_b$ equally spaced points according to arc-length, and each superior-inferior pair is connected by a straight segment. This simple geometric construction has been previously used to compute the callosal centerline and generate its corresponding diffusion map profile plots~\citep{park_relationships_2011,rittner_analysis_2013}.

    \item \textbf{Laplace streamlines:} the Laplace equation method for CC thickness estimation, originally introduced in~\citet{adamson_thickness_2011}, later incorporated into the software pipeline described in~\citet{adamson_software_2014} and publicly available on NITRC\footnote{https://www.nitrc.org/projects/ccsegthickness}, was used to compute $N_b$ streamlines across the CC. Because the method expects a discrete voxel grid, the superior and inferior splines were discretized using a coordinate grid upsampled by a factor of 20, ensuring smooth integer-valued boundary coordinates. Each resulting Laplace streamline was subsequently fitted with a spline without smoothing.
    
\end{itemize}

Each transverse curve was uniformly resampled at $N_t$ points, yielding a total of $N_t \times N_b$ parameterization points (Fig.~\ref{fig:CC_param}b). For the Laplace streamlines, the points were then downsampled to match the original image resolution.

\subsubsection{Sampling and Analysis}
\label{sec:sampling_analysis}

The parameterization points were used to sample the diffusion-derived scalar maps. For each individual, the map values were interpolated at every point using bilinear interpolation, producing standardized $N_t \times N_b$ matrices for each scalar map (Figs.~\ref{fig:CC_param}c and~\ref{fig:CC_param}d). Because it establishes point-to-point correspondence across individuals, this representation enabled pointwise comparisons independent of differences in CC shape or size.

\begin{figure*}[b!]
\centering
\includegraphics[width=\textwidth]{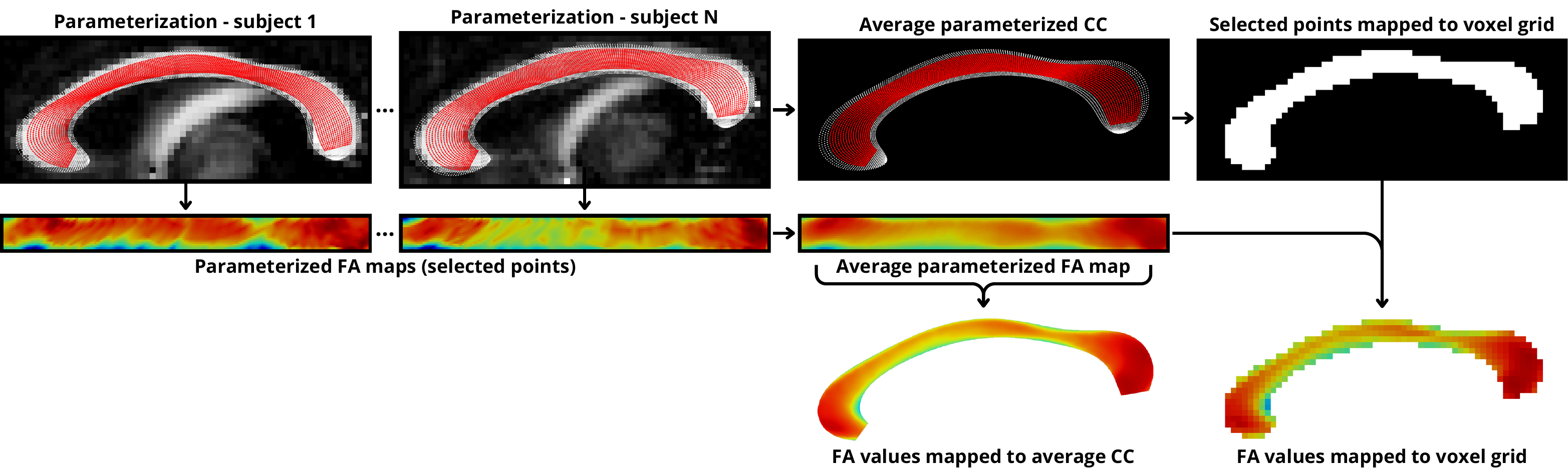}
\caption{\rev{Overview of the analysis and visualization steps using the parameterization method. Discarded parameterization points are represented in white and selected points in red. Parameterized measures can be mapped to a higher resolution CC template for visualization purposes (third column). For the analyses, measures are mapped to the voxel grid (fouth column).}}
\label{fig:CC_param_vis}
\end{figure*}

To mitigate partial-volume artifacts and segmentation inaccuracies, which are more likely to occur near the CC boundaries, the method allowed for the exclusion of the outermost parameterization points. The retained points restrict subsequent analyses to the more robust central portion of the CC (Fig.~\ref{fig:CC_param_vis}).

% -------------------------------------------------
% -------------------------------------------------

The spatial coordinates of the parameterization points across all individuals could be averaged to generate a template CC representation. Results derived in the parameterized space could then be mapped onto this template for a more intuitive visualization (Fig.~\ref{fig:CC_param_vis}). Alternatively, any external template CC mask (e.g., obtained by registering and combining individual CC masks) could be parameterized and used for mapping.

\rev{The parameterization produces an irregular sampling grid, with a higher concentration of sampling points in some regions, particularly at the CC extremities. Averaging across all points would therefore bias global measures toward oversampled regions. In addition, because the grid is denser than the native voxel grid in many regions and neighboring points draw on the same voxel neighborhood, many measurements are redundant and non-independent, which would underestimate variance in subsequent statistical analyses. To address both issues, values were mapped to the voxel grid of the CC template before computing descriptive statistics or statistical tests.}

\rev{The final template mask was defined as the region enclosed by the boundary of the selected points. Each value of interest in the $N_t \times N_b$ matrix was assigned to the corresponding parameterized template coordinates and, thus, to the nearest voxel. The values of points falling in the same voxel were averaged (Fig.{~\ref{fig:CC_param_vis}}).} An implementation of the parameterization method is publicly available at \url{https://github.com/MICLab-Unicamp/Midsagittal-CC-Parameterization} (under preparation).

\subsection{Datasets and Preprocessing}
\label{sec:datasets}

We used three datasets in this work. The test-retest dataset from the Human Connectome Project (HCP) consortium~\citep{van_essen_human_2012} was used to determine an optimal configuration of the parameterization method. This dataset, referred to as HCP-TR, includes publicly available T1w and diffusion data from 44 healthy participants, each scanned on two separate days. The T1w images had isotropic voxels of $0.7$ mm, while the diffusion images had isotropic voxels of $1.25$ mm and included three diffusion-weighted shells ($b=1000$, $2000$, and $3000$ s/mm$^2$ with 90 directions per shell), in addition to 18 $b=0$ s/mm$^2$ volumes. We used the minimally preprocessed data released by the HCP~\citep{glasser_minimal_2013}, with diffusion data aligned to each native structural space following intensity normalization, eddy-current and motion correction, and gradient-distortion correction.

Subsequently, we performed two example applications of DTI-based group analyses using selected methodologies. First, we replicated an earlier study by \textcite{sheriff_comparison_2024} that compared DTI scalar metrics in younger versus older healthy adults. From the IXI dataset\footnote{https://brain-development.org/ixi-dataset/}, we selected diffusion data acquired at Hammersmith Hospital. Images were obtained on a Philips Intera 3-T scanner with a voxel size of $1.75 \times 1.75 \times 2.0$ mm$^3$, 15 diffusion directions at $b=1000$ s/mm$^2$, and one $b=0$ volume. Following the demographic criteria adopted by \citet{sheriff_comparison_2024}, subjects aged 25-35 years were assigned to the younger adult (YA) group, and those aged 65 years or older to the older adult (OA) group.

Second, we compared DTI scalar metrics in individuals with AD to cognitively normal (CN) adults using the Alzheimer's Disease Neuroimaging Initiative (ADNI) dataset\footnote{https://adni.loni.usc.edu/}. For each participant in the ADNI3 and ADNI4 studies classified as AD or CN, we selected the earliest diffusion scan labeled ``Axial DTI'' in the ``Image Description'' field that met the following criteria: isotropic voxels of $2$ mm, 48 diffusion directions at $b=1000$ s/mm$^2$, and seven $b=0$ volumes. The groups were then matched for age and sex.

For both IXI and ADNI datasets, preprocessing included denoising, eddy-current and motion correction, rigid alignment to the MNI152 template (translation and rotation only), resampling to $1.25$ mm isotropic resolution, and brain extraction, using tools from MRtrix3~\citep{tournier_mrtrix_2012} and FSL~\citep{jenkinson_fsl_2012}. Rigid alignment was performed to reduce variability in midsagittal slice selection due to differences in head positioning, and resampling ensured consistency with the resolution of the HCP-TR dataset.

In all datasets, diffusion tensor reconstruction was performed using the DIPY package~\citep{garyfallidis_dipy_2014}. Scalar maps were generated for FA, MD, RD, and AxD. The midsagittal slice of each individual was identified using the diffusion characteristics of the interhemispheric fissure, a large region of cerebrospinal fluid (CSF, low FA) and adjacent white matter structures such as the CC (high FA). After removing slices with a small brain cross-sectional area, the midsagittal slice (containing mostly CSF) was defined as the slice with the lowest mean FA value~\citep{freitas_watershed-based_2012}.

The midsagittal CC segmentation was obtained by combining the automated TractSeg bundle segmentation~\citep{wasserthal_tractseg_2018} with targeted post-processing. \rev{TractSeg is a deep-learning method that segments many WM tracts, including the CC, using fiber orientation distribution peaks estimated from the dMRI data. The available tool\footnote{https://github.com/MIC-DKFZ/TractSeg} encapsulates most of the processing steps, so that preprocessed dMRI data can be used directly as input.} The binary CC bundle mask obtained with TractSeg was then restricted to the midsagittal slice and its two immediate neighbors. A connected-component analysis was performed, and the component whose centroid was closest to the center of the midsagittal brain mask was selected as the CC. The final midsagittal CC mask was obtained by repeating the connected-component analysis on the midsagittal slice alone and retaining the largest component. The masks were visually inspected and manually corrected as required.

For comparison purposes, we also applied the standard TBSS pipeline~\citep{smith_tract-based_2006} using the FSL toolbox~\citep{jenkinson_fsl_2012}. Briefly, individual FA volumes were nonlinearly registered to the FMRIB58 FA template and transformed into MNI152 space. A mean FA image was computed and thinned to produce the mean FA skeleton using a threshold of 0.2, after which each FA dataset was projected onto the skeleton. The same registration and projection steps were applied to the MD, RD, and AxD maps. Only the portion of the skeleton corresponding to the midsagittal CC was considered.

To obtain a representative midsagittal CC mask in the same space as the TBSS skeleton, we applied the TBSS registration transforms to the three-slice CC masks and averaged them across individuals. We then identified the midsagittal slice of the mean FA image, extracted the corresponding slice from the averaged CC mask, and binarized it using a threshold of $0.5$. Unless otherwise stated, this groupwise midsagittal CC mask was used as the template for mapping the parameterization results, enabling direct comparison against the TBSS-derived results.

\subsection{Configuration and Repeatability Experiments}

To determine an appropriate configuration of the parameterization method, we evaluated different configurations by varying both the number of sampling points and the definition of transverse curves. Our objective was to select parameters that (i) provide adequate spatial coverage of the midsagittal CC and (ii) yield stable measurements across test-retest acquisitions.

First, we used the entire HCP-TR cohort to calculate the average length of the centerline ($L$), the average length of the transverse curves ($T$), and their integer ratio ($r = \text{round}(L/T)$). To preserve the elongated geometry of the CC, we defined the number of boundary samples ($N_b$) proportionally to the number of transverse samples ($N_t$) as $N_b = r \cdot N_t$.

To explore the effect of sampling density, we tested values of $N_t$ from 5 to 41 in steps of 2. Odd values were used to ensure the presence of a centerline. A minimum of 5 points was chosen because lower values would bias the sampling toward the boundary regions, which are more susceptible to partial volume effects and segmentation inaccuracies.

After selecting values for $N_t$ and $N_b$, we assessed the effect of excluding sampling points near the CC boundaries. We first evaluated measurements along a central portion of the centerline and compared them with TBSS-derived results. We then progressively added pairs of longitudinal curves, followed by transverse curves, until the final sampling configuration was established.

Because one of the goals of the parameterization method is to capture a large portion of the midsagittal CC, we evaluated the sampling density and CC coverage. Additionally, we assessed the impact of the configurations on two repeatability measures: (i) the absolute error relative to the mean ($\varepsilon$) and (ii) the coefficient of variation (CoV). \rev{Using the final sampling configuration, we evaluated how these repeatability measures change with the input CC mask, by perturbing the original masks and by replacing them with masks from a different segmentation method.}

\subsubsection{Sampling Density and Coverage}

The sampling density was quantified by computing the number of parameterization points within each voxel. For each individual, a points-per-voxel map was generated and subsequently sampled using the parameterization grid. This process assigned a local sampling density value to each point in the $N_t \times N_b$ grid (analogous to how DTI scalar maps are sampled), enabling the calculation of a mean sampling density map across individuals.

To characterize the behavior of different parameterization configurations, we computed descriptive statistics (mean, minimum, and maximum) of the sampling density \rev{after projecting the density values onto the discrete voxel grid of the CC template (as described in Section{~\ref{sec:sampling_analysis}}).}
Furthermore, sparse configurations may fail to place sampling points in every voxel of the CC masks. To quantify this, we calculated the CC coverage for each individual, defined as the percentage of voxels within the individual's binary CC mask that contained at least one parameterization point.

\subsubsection{Repeatability Metrics}

We used two repeatability metrics:

\begin{itemize}

    \item The \textbf{absolute error relative to the mean} ($\varepsilon$)~\citep{jovicich_multisite_2014} was computed for each pair of test-retest measurements, indicating how much a single individual’s value changed between scans:

    $$\varepsilon = 100 \cdot \frac{|retest-test|}{(retest+test)/2}$$

    \item The \textbf{coefficient of variation} (CoV)~\citep{marenco_regional_2006} was computed across the entire group of individuals, indicating how large the variation (within-subject standard deviation, $SD_w$) is compared to the mean of all measurements ($\bar{x}$):

    $$CoV = 100 \cdot \frac{SD_w}{\bar{x}}$$
    
\end{itemize}

Both the error and the CoV were computed for every parameterization point and DTI map. Because the error is calculated per individual, we obtained the average results across individuals. For interpretability, we also computed the overall mean of each metric for every tested configuration (per map and then aggregated across maps). As in the sampling density analysis, the metric values were projected from the parameterization grid onto the CC template and averaged voxelwise before calculating global statistics.

\subsubsection{Sensitivity to Input Mask}
\label{sec:sensitivity_test_retest}

\rev{To evaluate how the parameterization depends on the input CC mask, the TractSeg-derived CC masks were perturbed in two ways: by eroding or dilating their boundaries, to assess the effect of over- and under-segmentation, and by replacing them with masks from a different segmentation method. The erosion and dilation used one iteration of a 4-connected structuring element, applied in-plane to each slice of the three-slice masks. For the alternative segmentation, volumetric CC masks were obtained with MRICloud~}\citep{mori_mricloud_2016}\rev{, a multi-atlas method that segments brain structures on T1w images. After computing the segmentation, the labels corresponding to the CC were selected and resampled from $0.7$ mm in the T1w to $1.25$ mm using the diffusion images as reference, with linear interpolation. The masks were then binarized at a threshold of 0.5 and reduced to the largest connected component.}

\rev{All masks were processed as the original TractSeg masks (Section{~\ref{sec:datasets}}) to obtain the midsagittal CC segmentations and the TBSS-derived CC templates. The masks were visually inspected and manually corrected as required. The repeatability analysis was performed using the final sampling configuration defined with the original TractSeg masks (Section{~\ref{sec:final_sampling_configuration}}), kept fixed so that differences reflect only the input mask, and sensitivity was assessed as the change in the repeatability metrics.}

\rev{Because each mask type yields a different template and therefore a different set of voxels, the analysis was repeated restricted to the voxels present in the templates of all mask types. This compares the same central voxels across mask types, isolating the effect of boundary changes from differences in template extent.}

\rev{Additionally, to assess the effect of localized differences in the mask, as opposed to the global boundary shifts introduced by erosion and dilation, we restricted the analysis to the subjects whose masks required correction, comparing the repeatability metrics obtained with the corrected and uncorrected versions of the same masks. As this analysis considered only parameterization results, without comparison to TBSS, the CC templates derived directly from the parameterization were used instead of the TBSS-derived templates. The uncorrected and corrected masks are shown in the Supplementary Material.}

\subsection{Example Applications}

To illustrate the proposed framework, we conducted two DTI-based group analyses using the IXI (YA vs. OA) and ADNI (AD vs. CN) datasets. In both cases, results obtained with the final parameterization were compared against the standard TBSS pipeline (as described in Section~\ref{sec:datasets}), the geometric parcellation method proposed by \citet{hofer_topography_2006} and a conventional VBM-style analysis.

To improve comparability between the geometric parcellation and the proposed parameterization, the parcellation analysis was restricted to the region defined by the parameterization. Specifically, midsagittal CC segmentations were first parcellated according to Hofer’s scheme using the inCCsight tool~\citep{caldeira_inccsight_2021}. Rather than averaging diffusion values across all voxels within each CC region, we considered only those voxels corresponding to the selected parameterization points. Results obtained using all voxels within each region are provided in the Supplementary Material.

For the VBM-style analysis, we used the DTI maps registered to the FMRIB58 FA template and transformed to the MNI152 space by TBSS. A threshold of $0.2$ was applied to the mean FA image to define the WM mask. The segmented DTI maps were smoothed with a Gaussian kernel with $4$ mm full-width at half-maximum (FWHM). Because VBM-style results may depend on the degree of smoothing~\citep{van_hecke_dti_2016}, additional analyses were performed without smoothing and with $2$, $6$, and $8$ mm FWHM smoothing (see Supplementary Material).

\rev{Voxelwise statistical analyses were performed using FSL’s randomise tool~}\citep{winkler_permutation_2014}\rev{, which fits a general linear model (GLM) at each voxel~}\citep{friston_statistical_1994}\rev{. Group differences were tested with two-sample unpaired $t$-tests, including sex as a nuisance regressor in the IXI analysis and both sex and age in the ADNI analysis, with 5,000 permutations and family-wise error (FWE) correction for multiple comparisons (corrected $p<0.05$).} Threshold-free cluster enhancement (TFCE) was applied for all methods except the geometric parcellation, for which voxelwise FWE correction was used.

\rev{For the parameterization-based analyses, values were mapped to the voxel grid of the CC template before testing, as described in Section{~\ref{sec:sampling_analysis}}. This mapping also provides the regular spatial neighborhood required by TFCE: on the irregular sampling grid, points are unevenly spaced in physical space and densely packed at the CC extremities, so the cluster extent accumulated by TFCE would reflect sampling density instead of spatial extent.} For TBSS and VBM-style analyses, randomise was applied to the full skeleton and WM volume, respectively, and the midsagittal slice was subsequently extracted.

\rev{To describe the magnitude of group differences independently of sample size, the unthresholded $t$-statistic maps were converted to Cohen’s $d$ effect-size maps~}\citep{cohen_statistical_2013}\rev{ and corrected for small-sample bias using Hedges’ factor, yielding Hedges’ $g$}~\citep{hedges1985}\rev{. To maintain consistency with the statistical tests, effect sizes were expressed relative to the GLM’s residual standard deviation after accounting for the covariates (sex for IXI; sex and age for ADNI), following the approach implemented in the \texttt{emmeans} R package~}\citep{lenth2026}\rev{. We report unthresholded $g$-maps, showing the full continuous effect-size topography across methods, and $g$-maps restricted to regions surviving FWE correction ($p<0.05$).}

\subsubsection{Quantitative Results}

\rev{Quantitative comparisons were performed using three complementary measures of spatial similarity and agreement between methods. First, the Dice coefficient~}\citep{dice1945measures}\rev{ was calculated between binarized FWE-significant masks to quantify the spatial overlap of significant clusters. Second, Pearson's correlation coefficient ($r$) was calculated between unthresholded Hedges' $g$ effect-size maps to quantify the linear association between effect sizes across spatial locations.
Finally, Lin's concordance correlation coefficient (CCC)~}\citep{lin_concordance_1989}\rev{ was used to assess agreement in both spatial patterns and effect-size magnitudes. Unlike Pearson's $r$, which measures linear association, CCC quantifies concordance relative to the line of identity ($y=x$), accounting for differences in the means and variances of the two maps. It is defined as $ \mathrm{CCC} = r C_b $, where $C_b$ is the bias correction factor, given by}

$$ C_b = \frac{2\sigma_1\sigma_2} {\sigma_1^2+\sigma_2^2+(\mu_1-\mu_2)^2} = \frac{2}{\nu+\nu^{-1}+u^2}\text{,} $$

\rev{with $\nu=\sigma_1/\sigma_2$ representing the ratio of standard deviations and $u=(\mu_1-\mu_2)/\sqrt{\sigma_1\sigma_2}$ representing the standardized difference in means. Values of $C_b$ closer to 1 indicate smaller discrepancies in location and scale, whereas lower values indicate greater disagreement. By combining this bias correction with Pearson's $r$, CCC accounts for differences in effect-size distributions that may not be reflected in correlation alone.}

\rev{Parcellation was excluded from these analyses because it produces region-wise rather than voxelwise results. Two comparison settings were considered: (1) pairwise comparisons among parameterization, TBSS, and VBM-style analyses, restricted to voxels within the TBSS skeleton mask; and (2) comparisons between parameterization and VBM-style analyses within the CC mask derived from the parameterization method.}

\subsubsection{Sensitivity to Input Mask}

\rev{We also assessed the impact of input mask variations on group-level statistical results. For each dataset (IXI and ADNI), the TractSeg-derived CC masks of each individual were eroded and dilated as described in Section{~\ref{sec:sensitivity_test_retest}}.
The statistical analyses with FSL's randomise were restricted to the intersection of the parameterization-derived CC templates obtained using the original, eroded, and dilated masks, ensuring a fair comparison and that differences in the results were not attributable to differences in the spatial extent of the analysis mask.
The resulting $t$-statistics were converted to Hedges' $g$ effect sizes, and the quantitative analyses described above (Dice coefficient, Pearson's correlation coefficient, and Lin's CCC) were used to compare the results across mask types. MRICloud-derived masks were not evaluated in this analysis because the T1w and diffusion images were not in the same space in both the IXI and ADNI datasets.}

% ===========================================================
\section{Results}
\label{sec:results}
% ===========================================================

\subsection{Configuration and Repeatability Experiments}

\subsubsection{Number of Sampling Points}
\label{sec:n_points}

Using straight lines, the average length of the centerline across individuals ($L$) was $70.04$ voxels ($87.55$ mm), while the average length of the transverse straight lines ($T$) was $7.56$ voxels ($9.45$ mm). This yielded an integer ratio $r = 9$, which resulted in $N_b$ ranging from $45$ to $369$ points across the tested configurations. For the Laplace streamlines, the length $L$ was $70.36$ voxels ($87.95$ mm), while the average length of the transverse Laplace streamlines ($T$) was $5.41$ voxels ($6.76$ mm), resulting in $r = 13$ and $N_b$ ranging from $65$ to $533$ points.

% -------------------------------------------------
\begin{figure*}[t!]
\centering
\includegraphics[width=0.6\textwidth]{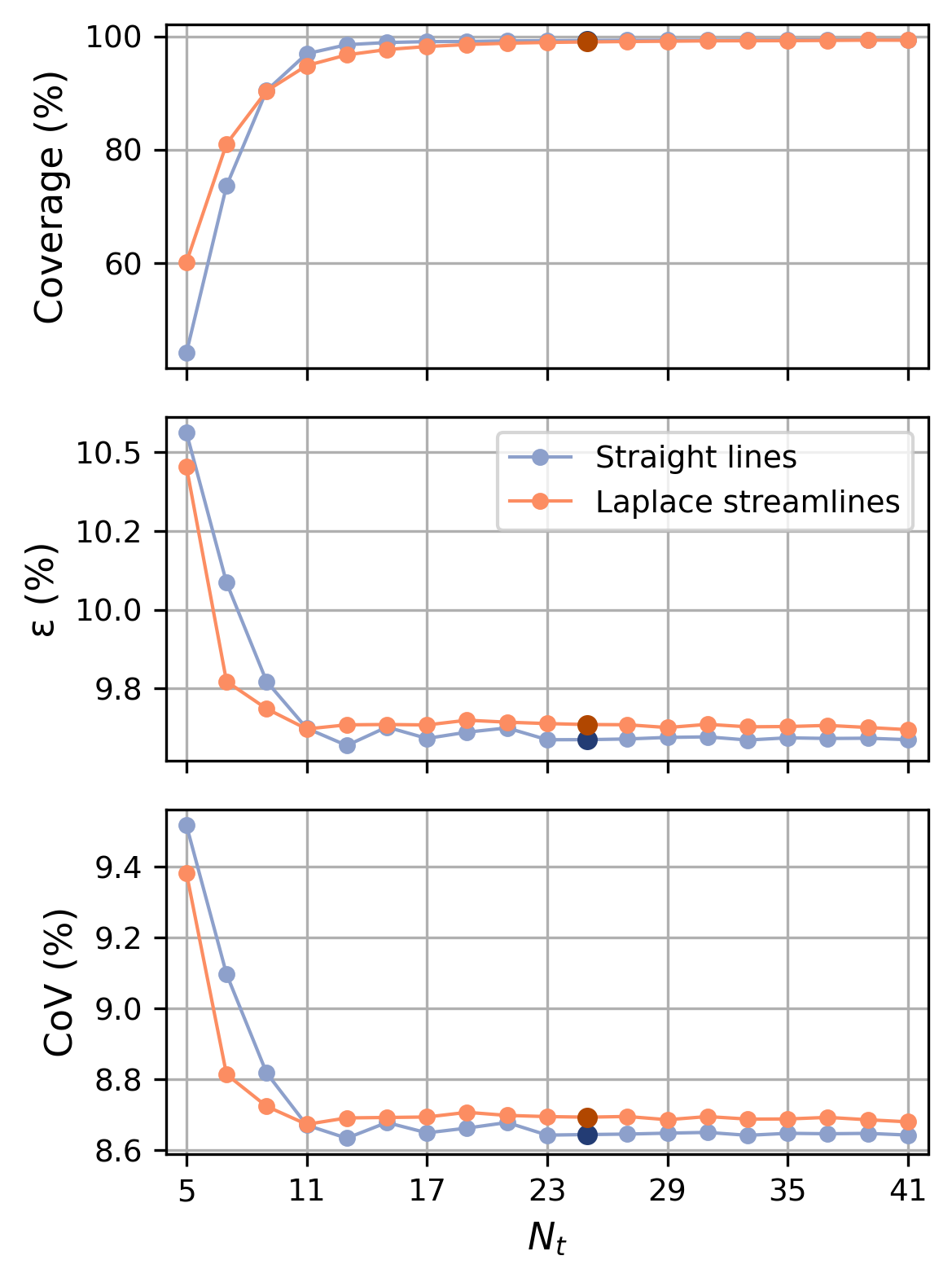}
\caption{Results for each $N_t$ value, using straight lines or Laplace streamlines. Coverage values were averaged across individuals, while repeatability metrics ($\varepsilon$, CoV) were averaged across individuals, DTI maps, and voxels. The darker points mark the selected configuration.}
\label{fig:npoints}
\end{figure*}
% -------------------------------------------------

In both cases, the analysis showed a rapid increase in sampling coverage and a simultaneous decrease in both error ($\varepsilon$) and CoV for the initial increments of sampling points (Fig.~\ref{fig:npoints}). The three metrics began to stabilize around $N_t = 11$, with the coverage of the Laplace streamlines exhibiting a less steep increase compared to the straight lines. Nevertheless, as more points were added, coverage continued to increase with progressively smaller gains.

While the repeatability metrics exhibited only minor fluctuations after $(N_t\times N_b) = (11\times 99)$, they showed improved stability for the straight lines after $(N_t\times N_b) = (23\times 207)$. Therefore, we selected $(N_t\times N_b) = (25\times 225)$ as its initial sampling configuration. For the Laplace streamlines, there was no clear point after $(N_t\times N_b) = (11\times 143)$ where stability improved. Based on the coverage results and considering the straight line configuration, we selected $(N_t\times N_b) = (25\times 325)$ as its initial sampling. The overall results using these configurations were:

\begin{itemize}

    \item \textbf{Straight lines:} coverage of $99.27\%$, $\varepsilon=9.59\%$, $\text{CoV}=8.64\%$, and a mean sampling density of $12.13$ points per voxel (ranging from $3.21$ to $55.89$ points per voxel).

    \item \textbf{Laplace streamlines:} coverage of $99.00\%$, $\varepsilon=9.64\%$, $\text{CoV}=8.69\%$, and a mean sampling density of $20.17$ points per voxel (ranging from $2.94$ to $49.52$ points per voxel).
    
\end{itemize}

\begin{figure*}[t!]
\centering
\includegraphics[width=0.9\textwidth]{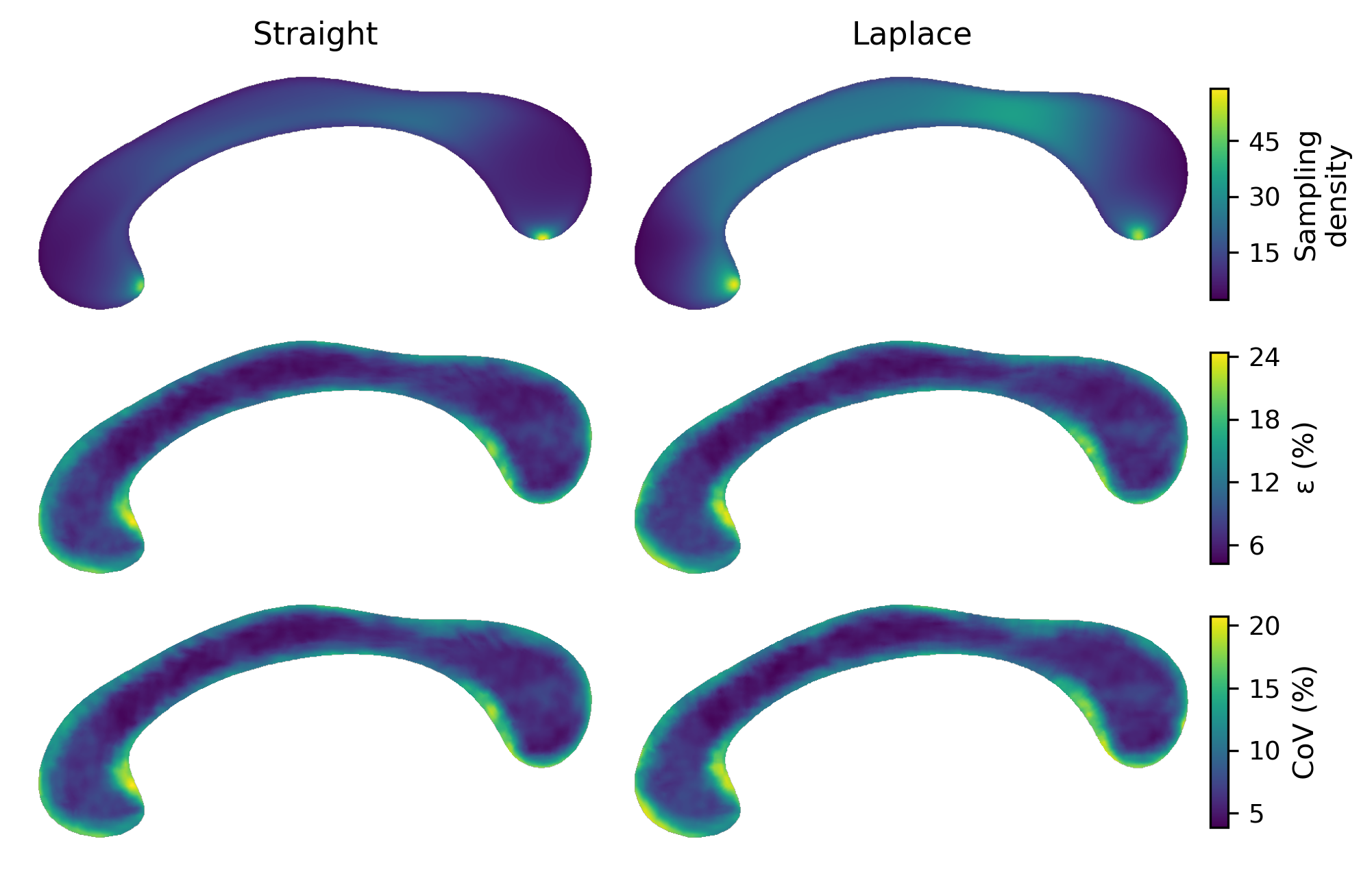}
\caption{Spatial distribution of sampling density, absolute error relative to the mean ($\varepsilon$), and coefficient of variation (CoV) on the CC template for the selected $(N_t\times N_b)$ configuration: $25\times 225$ (straight lines, first column) and $25\times 325$ (Laplace streamlines, second column). Repeatability metrics ($\varepsilon$, CoV) were averaged across individuals and DTI maps.}
\label{fig:all_pts_distribution}
\end{figure*}

The pointwise distribution of the metrics along the CC template indicated higher sampling densities near the endpoints (Fig.~\ref{fig:all_pts_distribution}). When using Laplace streamlines, the sampling density was also higher in the central portion of the CC. Overall, regions with lower repeatability tended to concentrate near the CC boundaries. A similar pattern was observed across DTI maps, but error and CoV values were particularly higher for RD, as well as for FA in specific areas near the inferior boundary (Supplementary Figs.~1 and~2).

% -------------------------------------------------
% -------------------------------------------------

\subsubsection{Final Sampling Configuration}
\label{sec:final_sampling_configuration}

\begin{figure*}[tb!]
\centering
  \begin{tabular}{@{} m{0.4cm} @{\hspace{0.2cm}} m{7cm} @{\hspace{0.7cm}} m{0.4cm} @{\hspace{0.2cm}} m{7cm} @{}}
  
    \small \textbf{(a)} & \fbox{ \includegraphics[width=7cm]{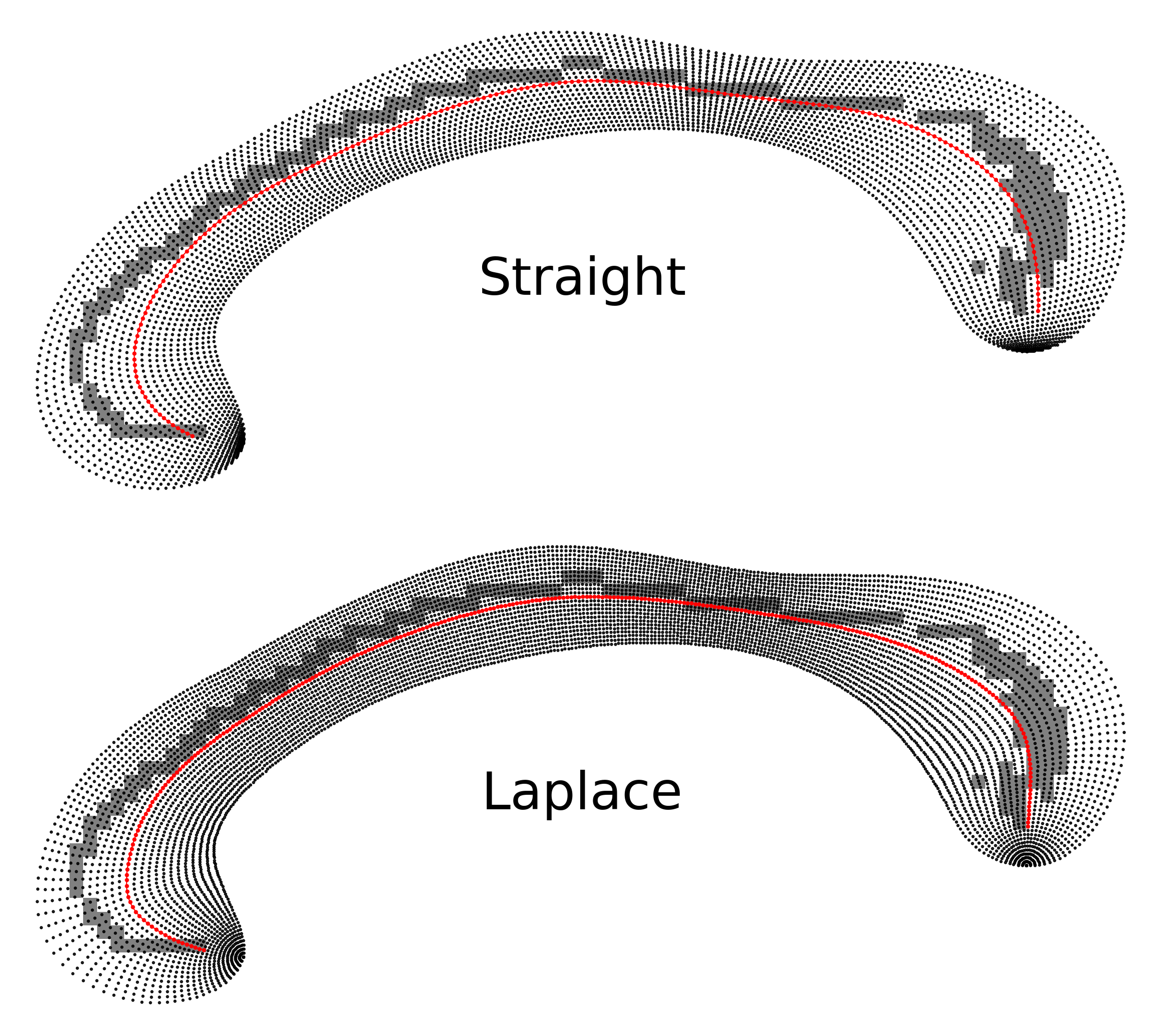} } & \small \textbf{(b)} & \fbox{ \includegraphics[width=7cm]{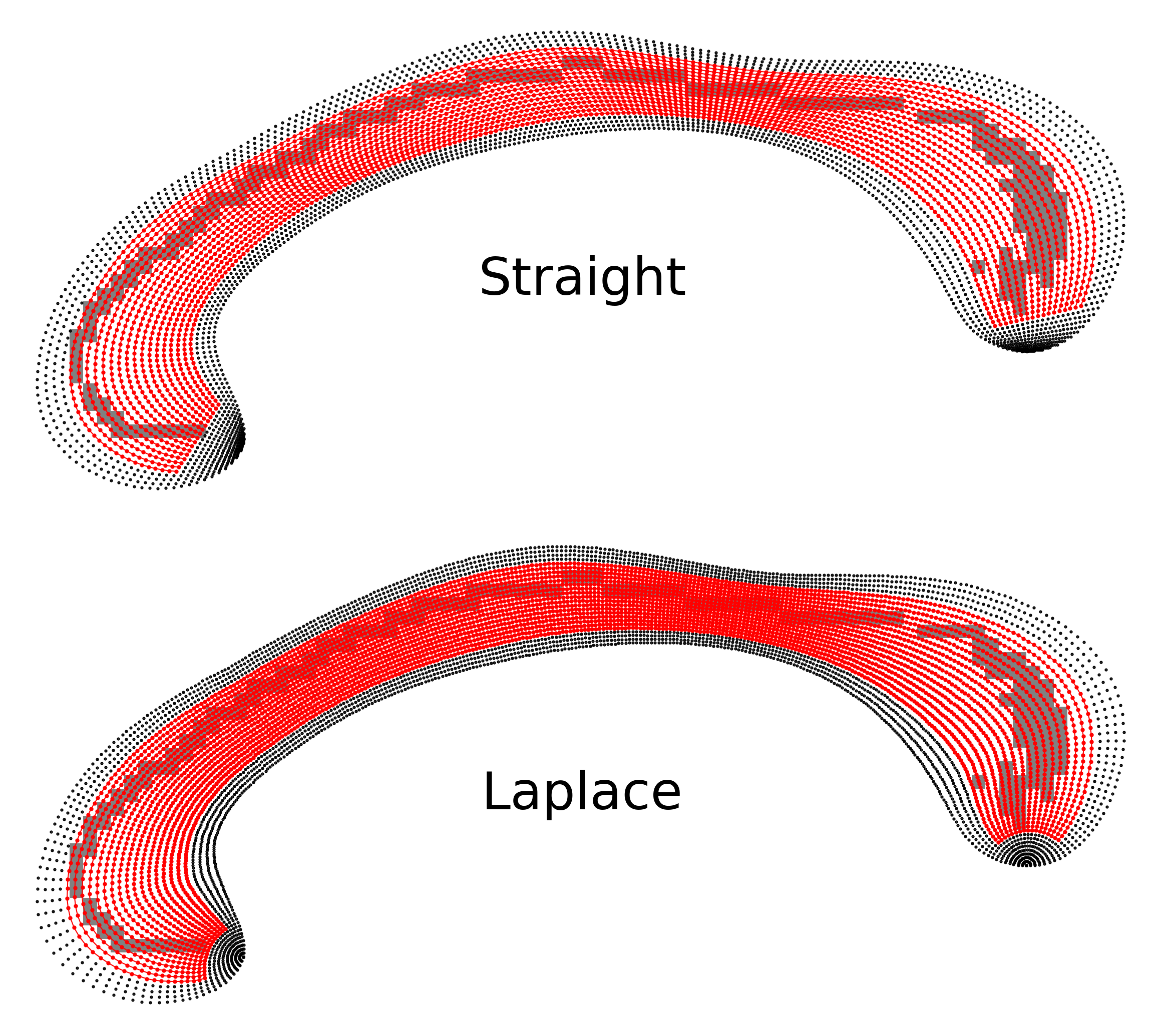} } \\
    
  \end{tabular}

\vspace{0.1cm}

\centering
  \begin{tabular}{@{} m{0.4cm} @{\hspace{0.2cm}} m{15.3cm} @{}}

    \small \textbf{(c)} & \fbox{ \includegraphics[width=15.3cm]{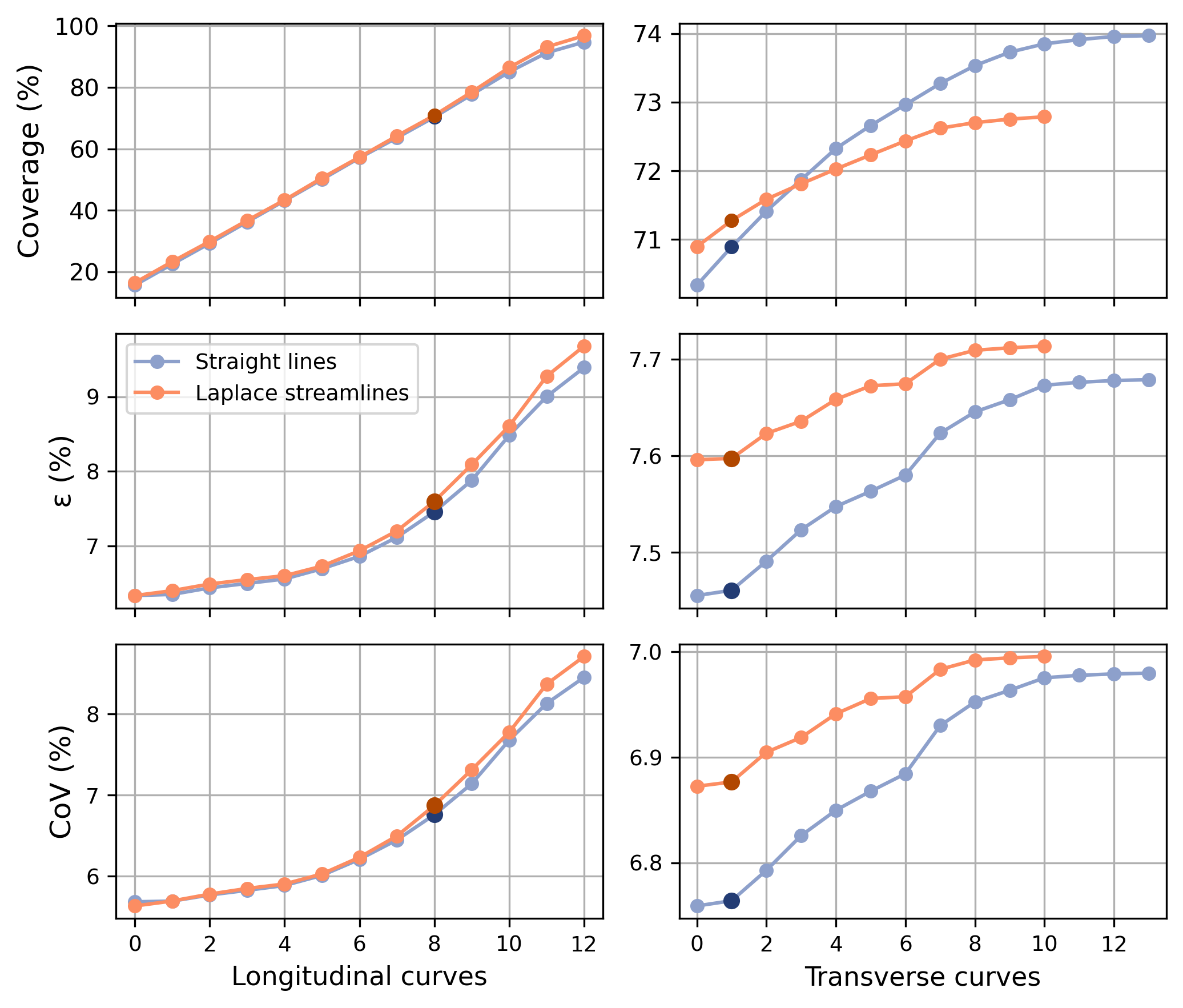} } \\
    
  \end{tabular}

\caption{\textbf{(a)} Portion of the centerline (red) extracted from the selected configuration (black dots), overlaid on the TBSS skeleton (gray area). \textbf{(b)} Final parameterization retaining only the central points (red). \textbf{(c)} Coverage and repeatability results obtained as pairs of longitudinal curves (left) and transverse curves (right) were progressively added. The darker points indicate the selected number of added pairs.}
\label{fig:rows_cols_addition}

\end{figure*}

Starting from the selected configurations, we disregarded points from each end of the centerline to make it approximately match the ends of the TBSS skeleton. We disregarded $13$ points for the straight lines and $10$ points for the Laplace streamlines, retaining the $199$ and $305$ central points, respectively (Fig.~\ref{fig:rows_cols_addition}a). Overall, these regions presented repeatability measures comparable to TBSS (Tables~\ref{tab:error} and~\ref{tab:cov}), with the parameterization centerline achieving higher repeatability for all maps except FA.

% -------------------------------------------------
% -------------------------------------------------

Next, we expanded the selection by progressively adding the closest pairs of superior and inferior longitudinal curves. Based on the repeatability metrics (Fig.~\ref{fig:rows_cols_addition}c, left), we incorporated eight pairs into the final configuration of each approach. We then applied the same incremental strategy in the transverse direction (Fig.~\ref{fig:rows_cols_addition}c, right), leading to the inclusion of one more pair of transverse curves. Therefore, from the initial grid of $(25\times 225)$ points, the final parameterization using straight lines retained the $(17\times 201)$ central points (Fig.~\ref{fig:rows_cols_addition}b). Similarly, from $(25\times 325)$ points, the final parameterization using Laplace streamlines kept $(17\times 307)$ points.

\begin{figure*}[b!]
\centering
\includegraphics[width=0.9\textwidth]{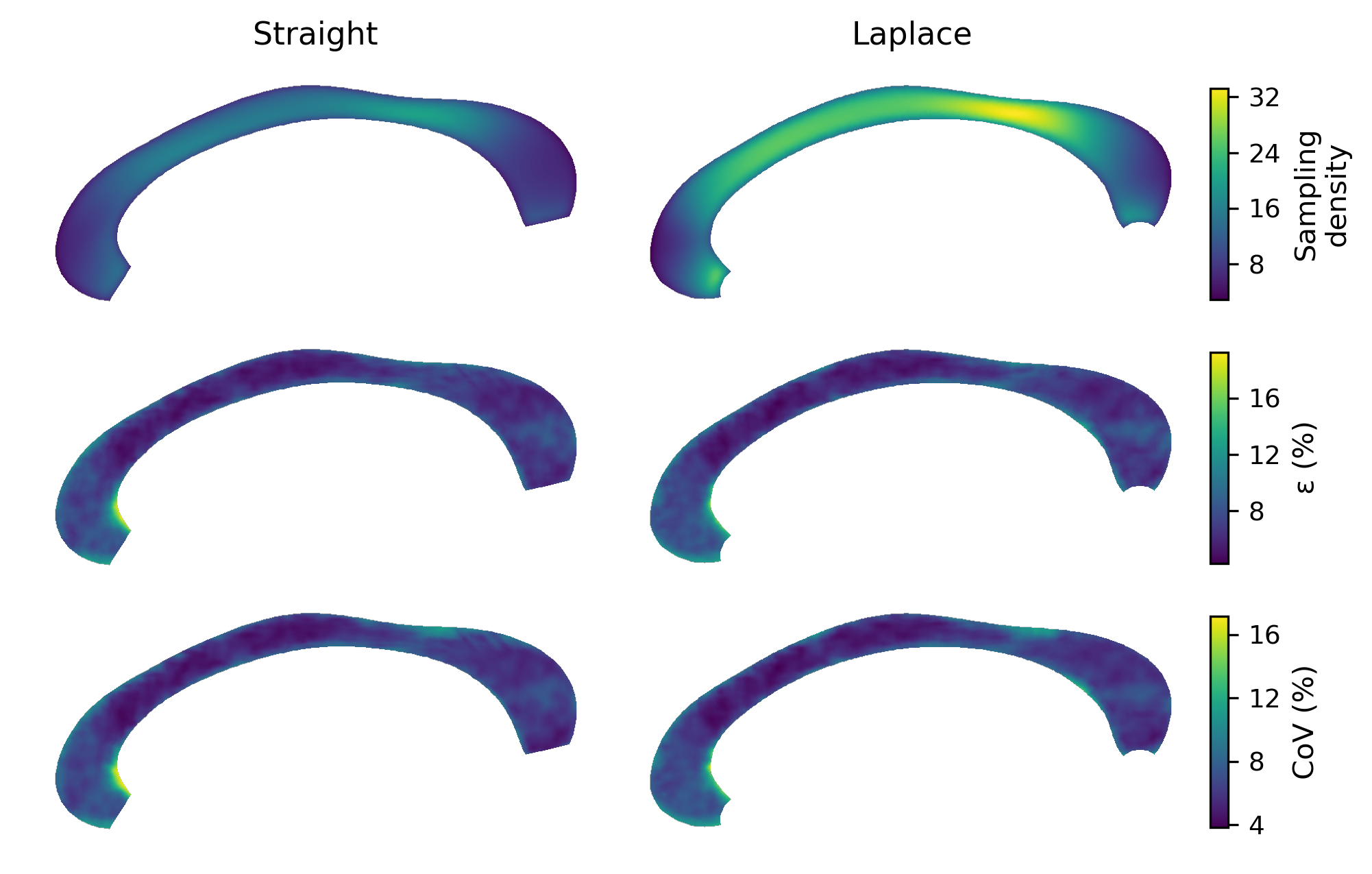}
\caption{Spatial distribution of sampling density, absolute error relative to the mean ($\varepsilon$), and coefficient of variation (CoV) on the CC template for the final configuration, after removal of the outermost points: $17\times 201$ (straight lines, first column) and $17\times 307$ (Laplace streamlines, second column). Repeatability metrics ($\varepsilon$, CoV) were averaged across individuals and DTI maps.}
\label{fig:selec_pts_distribution}
\end{figure*}

In both approaches, the final configuration constrained the analysis to regions of higher repeatability (Fig.~\ref{fig:selec_pts_distribution}, Supplementary Figs.~1 and~2). Although these configurations exhibited lower repeatability than the centerline alone or TBSS (Tables~\ref{tab:error} and~\ref{tab:cov}), they cover a greater portion of the CC, with a mean coverage of $70.89\%$ using the straight lines and $71.28\%$ using the Laplace streamlines.

% -------------------------------------------------
% -------------------------------------------------

Results were very similar for the straight-line and Laplace-based strategies (Tables~\ref{tab:error} and~\ref{tab:cov}). However, the Laplace method showed greater variation in sampling density (Fig.~\ref{fig:selec_pts_distribution}) and required more than 100 additional transverse curves, resulting in over 1,700 extra parameterization points. It was also more computationally demanding. After ten repetitions on an Intel Core i9-13900KF, the straight-line strategy with $(25\times 225)$ points required on average $13.8 \pm 0.6$ s to process the entire HCP-TR dataset using a single core, whereas the Laplace method required $2281.7 \pm 7.1$ s with the same configuration. Using $(25\times 325)$ points, the Laplace method required $3283.3 \pm 7.3$ s. Given the comparable repeatability but higher computational cost and sampling complexity, the straight-line approach was selected as the final parameterization configuration.

% -------------------------------------------------
\begin{table*}[t!]
\centering
\caption{Mean $\pm$ standard deviation of the absolute error relative to the mean ($\varepsilon$) for different methods and DTI maps.}
\begin{tabular}{lcccccc}
\toprule
& TBSS 
& \multicolumn{2}{c}{Parameterization (straight)} 
& \multicolumn{2}{c}{Parameterization (Laplace)} \\
\cmidrule(lr){3-4} \cmidrule(lr){5-6}
& 
& Centerline & Final configuration 
& Centerline & Final configuration \\
\midrule
FA & 2.463 $\pm$ 0.579 & 3.223 $\pm$ 0.642 & 4.730 $\pm$ 2.781 & 3.319 $\pm$ 0.656 & 4.823 $\pm$ 2.607 \\
MD & 5.562 $\pm$ 0.911 & 5.441 $\pm$ 0.959 & 6.126 $\pm$ 1.450 & 5.388 $\pm$ 1.004 & 6.204 $\pm$ 1.434 \\
RD & 13.186 $\pm$ 3.973 & 12.011 $\pm$ 3.160 & 13.901 $\pm$ 3.347 & 12.038 $\pm$ 3.232 & 14.251 $\pm$ 3.525 \\
AxD & 4.761 $\pm$ 1.136 & 4.650 $\pm$ 1.010 & 5.086 $\pm$ 1.217 & 4.578 $\pm$ 1.016 & 5.110 $\pm$ 1.216 \\
\bottomrule
\end{tabular}
\label{tab:error}
\end{table*}
% -------------------------------------------------

% -------------------------------------------------
\begin{table*}[t!]
\centering
\caption{Mean $\pm$ standard deviation of the coefficient of variation (CoV) for different methods and DTI maps.}
\begin{tabular}{lcccccc}
\toprule
& TBSS 
& \multicolumn{2}{c}{Parameterization (straight)} 
& \multicolumn{2}{c}{Parameterization (Laplace)} \\
\cmidrule(lr){3-4} \cmidrule(lr){5-6}
& 
& Centerline & Final configuration 
& Centerline & Final configuration \\
\midrule
FA & 2.209 $\pm$ 0.500 & 2.903 $\pm$ 0.553 & 4.235 $\pm$ 2.360 & 2.964 $\pm$ 0.551 & 4.335 $\pm$ 2.275 \\
MD & 5.030 $\pm$ 0.825 & 4.947 $\pm$ 0.844 & 5.626 $\pm$ 1.426 & 4.878 $\pm$ 0.868 & 5.686 $\pm$ 1.414 \\
RD & 11.567 $\pm$ 3.150 & 10.751 $\pm$ 2.698 & 12.636 $\pm$ 3.173 & 10.631 $\pm$ 2.620 & 12.906 $\pm$ 3.353 \\
AxD & 4.249 $\pm$ 1.061 & 4.140 $\pm$ 0.845 & 4.559 $\pm$ 1.119 & 4.063 $\pm$ 0.879 & 4.579 $\pm$ 1.131 \\
\bottomrule
\end{tabular}
\label{tab:cov}
\end{table*}
% -------------------------------------------------

\subsubsection{Sensitivity to Input Mask}

\rev{The spatial extension of the TBSS-derived CC templates varied by mask type, decreasing from the dilated to the original, MRICloud, and eroded masks (Supplementary Fig.~3).
In particular, the MRICloud and TractSeg-eroded templates were very similar in extension.}

\rev{Because the number of parameterization points was fixed, larger templates resulted in lower sampling density: the dilated masks presented the lowest values, followed by the original, MRICloud, and eroded masks (Supplementary Fig.~4). A similar order was observed for the repeatability measures, with the dilated masks showing the highest $\varepsilon$ and CoV, particularly at the most anterior portion of the inferior boundary. Compared with the original masks, the mean $\varepsilon$ and CoV (averaged across DTI maps) showed relative increases of $21\%$ for the dilated masks and relative decreases of $10\%$ for the eroded masks, whereas the MRICloud masks presented practically the same values (relative decreases of only $2\%$ and $1\%$, respectively) (Table{~\ref{tab:input_summary}}; see Supplementary Tables~1 and~2 for results per DTI map).}

% -------------------------------------------------
\begin{table}[t!]
\centering
\caption{\rev{Mean $\varepsilon$ and CoV for each type of input mask, considering all template voxels and only the voxels common to all templates. The metrics were averaged across DTI maps. TractSeg: original TractSeg. TractSeg-dil: dilated TractSeg. TractSeg-ero: eroded TractSeg.}}
\label{tab:input_summary}
\begin{tabular}{lcccc}
\toprule
 & \multicolumn{2}{c}{All voxels} & \multicolumn{2}{c}{Common voxels} \\
\cmidrule(lr){2-3}\cmidrule(lr){4-5}
Mask & $\varepsilon$ & CoV & $\varepsilon$ & CoV \\
\midrule
TractSeg      & 7.46 & 6.76 & 6.94 & 6.28 \\
TractSeg-dil  & 9.00 & 8.19 & 7.12 & 6.49 \\
TractSeg-ero  & 6.73 & 6.09 & 6.72 & 6.08 \\
MRICloud      & 7.31 & 6.71 & 7.19 & 6.59 \\
\bottomrule
\end{tabular}
\end{table}
% -------------------------------------------------

\begin{figure*}[tb!]
\centering
\includegraphics[width=0.8\textwidth]{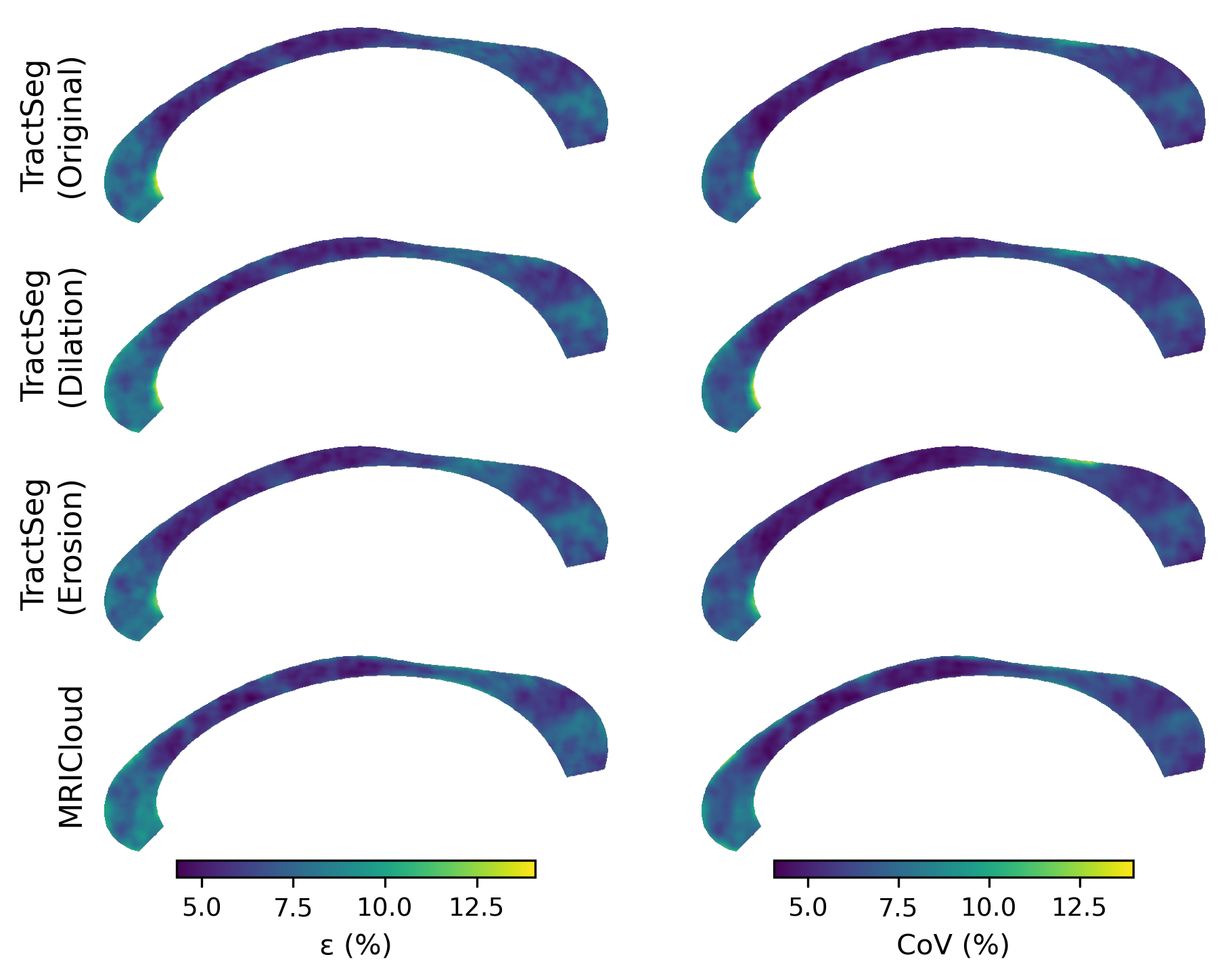}
\caption{\rev{Spatial distribution of sampling density, absolute error relative to the mean ($\varepsilon$), and coefficient of variation (CoV) on the CC template for each type of input mask, considering only the voxels common to all templates. Repeatability metrics ($\varepsilon$, CoV) were averaged across individuals and DTI maps.}}
\label{fig:selec_pts_distribution_input_common}
\end{figure*}

\rev{When the analysis was restricted to the voxels present in the templates of all mask types, the differences decreased markedly: the range of $\varepsilon$ fell from 2.27 to 0.47 percentage points, and that of CoV from 2.10 to 0.51 (Table{~\ref{tab:input_summary}}; see Supplementary Tables~3 and~4 for results per DTI map).
This was mainly due to the dilated masks, for which the relative increases in $\varepsilon$ and CoV compared with the original masks fell from $21\%$ to $3\%$ in both measures.
The eroded masks remained lower than the original masks (relative decreases of $3\%$ in both measures), with values practically unchanged relative to the analysis with all voxels.
In contrast to the full analysis, in which MRICloud was slightly lower than the original masks, the MRICloud masks showed relative increases of $4\%$ and $5\%$ in the common voxels and presented the highest values among mask types.
The most noticeable regional difference remained the higher values at the most anterior portion of the inferior boundary (Fig.{~\ref{fig:selec_pts_distribution_input_common}}).}

% -------------------------------------------------
% -------------------------------------------------

% -------------------------------------------------
\begin{table}[t!]
\centering
\caption{\rev{Comparison of corrected and uncorrected MRICloud masks for the four individuals in whom correction was required. Values are mean $\pm$ standard deviation of $\varepsilon$ and CoV per DTI map using parameterization-derived CC templates.}}
\begin{tabular}{lcccc}
\toprule
& \multicolumn{2}{c}{$\varepsilon$} & \multicolumn{2}{c}{CoV} \\
\cmidrule(lr){2-3} \cmidrule(lr){4-5}
& Corrected & Uncorrected & Corrected & Uncorrected \\
\midrule
FA
& $4.088 \pm 2.918$
& $4.467 \pm 3.846$
& $3.536 \pm 2.746$
& $3.792 \pm 3.297$ \\
MD
& $4.799 \pm 2.263$
& $4.802 \pm 2.481$
& $4.101 \pm 2.059$
& $4.089 \pm 2.113$ \\
RD
& $12.198 \pm 6.229$
& $12.260 \pm 6.427$
& $10.527 \pm 5.762$
& $10.462 \pm 5.546$ \\
AxD
& $4.367 \pm 2.220$
& $4.671 \pm 2.467$
& $3.650 \pm 1.937$
& $3.915 \pm 2.129$ \\
\bottomrule
\end{tabular}
\label{tab:corrected_uncorrected}
\end{table}
% -------------------------------------------------

\rev{Manual correction was only performed for four individuals using MRICloud (Supplementary Fig.~5) and the comparison between corrected and uncorrected masks was therefore restricted to them. Compared with the uncorrected masks, correction slightly reduced $\varepsilon$ and CoV for FA (relative decreases of $8.5\%$ and $6.8\%$, respectively) and AxD ($6.5\%$ and $6.8\%$), whereas MD and RD were essentially unchanged (relative changes below $1\%$, with opposite directions for $\varepsilon$ and CoV) (Table{~\ref{tab:corrected_uncorrected}}).}

\subsection{Example Applications}

\subsubsection{Younger \textit{vs}. Older Adults}

The final sample from the IXI dataset comprised 34 YA participants (14 females, mean ± standard deviation age: $29.6 \pm 3.2$ years) and 30 OA participants (18 females, $70.8 \pm 4.0$ years).

The average length of the centerline ($L$) was $71.72$ voxels ($89.65$ mm), while the average length of the transverse straight lines ($T$) was $7.91$ voxels ($9.89$ mm). These results yielded a ratio $r = 9$, similar to that observed in the HCP-TR dataset (see Section~\ref{sec:n_points}). The coverage of the initial sampling configuration ($(25\times 225)$ points) was $99.35\%$, whereas the coverage of the final sampling configuration ($(17\times 201)$ points) was $70.95\%$.

\begin{figure}[tb!]
  \centering
  \begin{tabular}{@{} m{0.8cm} @{} m{17cm} @{}}
    \small \textbf{(a)} & \includegraphics[width=17cm]{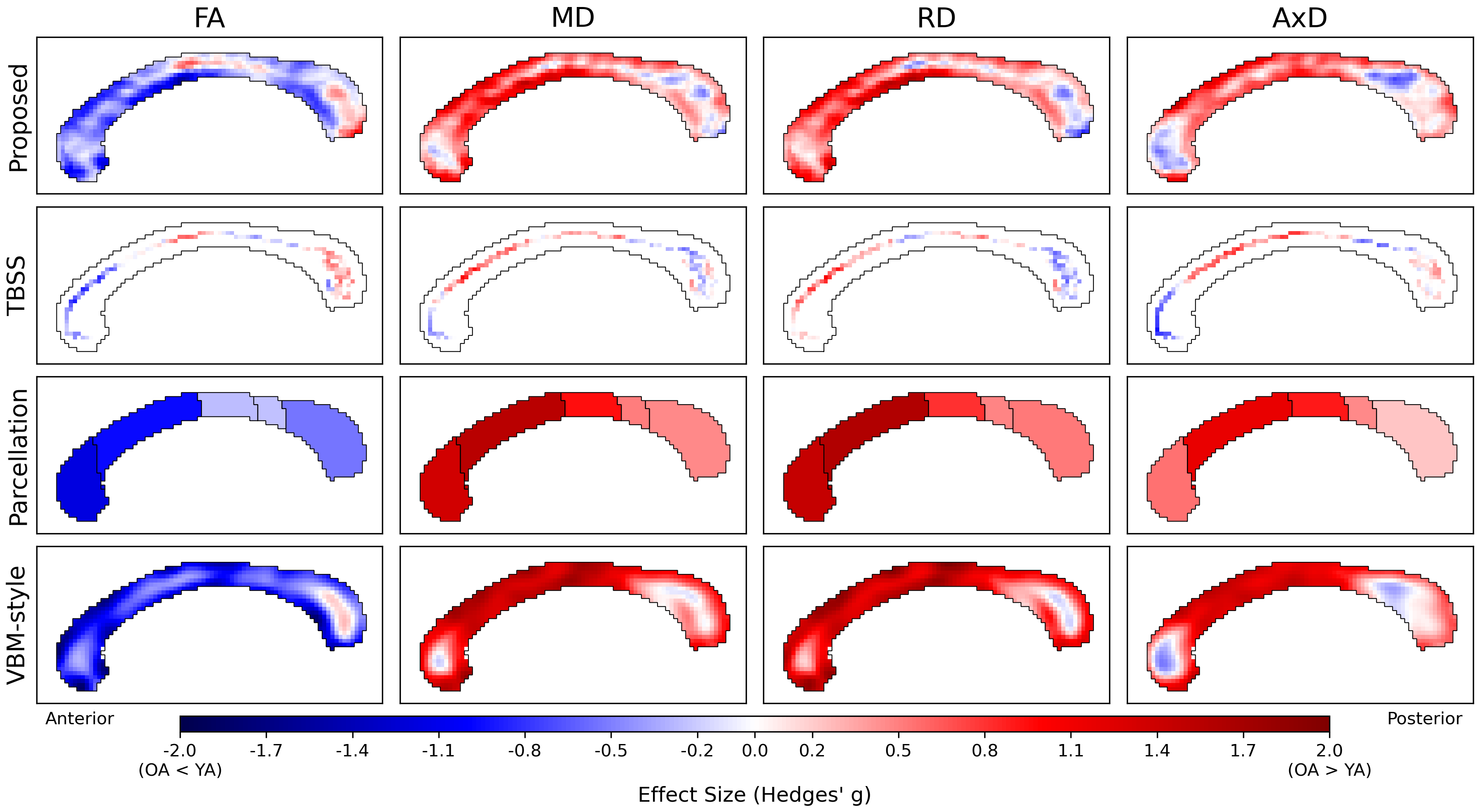} \\
    \small \textbf{(b)} & \includegraphics[width=17cm]{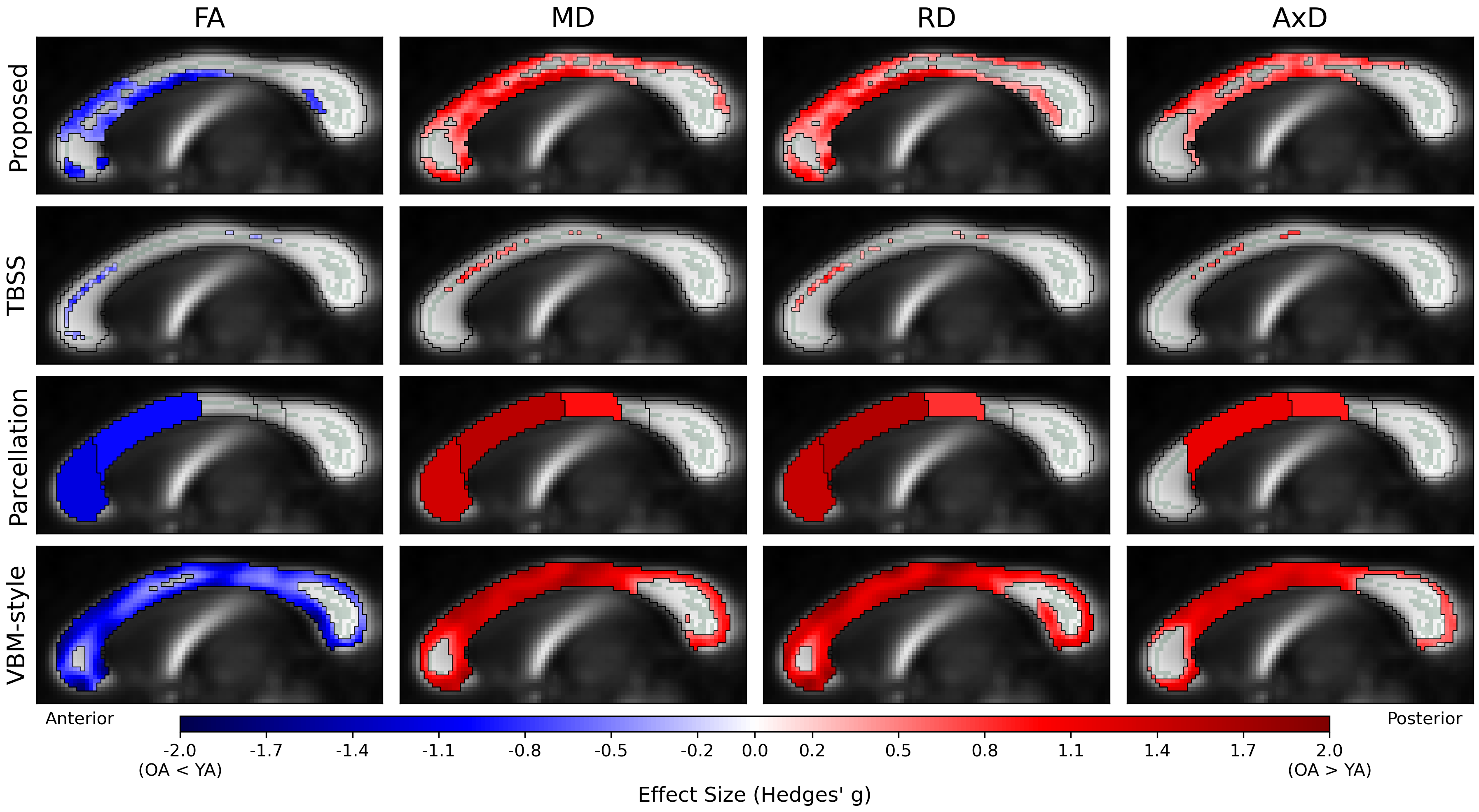}
  \end{tabular}
\caption{\rev{Group differences between younger adults (YA) and older adults (OA) in the IXI dataset, shown as Hedges’ $g$ maps. \textbf{(a)} Unthresholded maps. \textbf{(b)} Maps restricted to regions surviving FWE-corrected $p<0.05$  over mean FA image. Results are organized by method (rows) and DTI map (columns). TBSS and VBM-style results are shown within the CC parameterization mask (black contour). The grayish-green region denotes the TBSS skeleton of the midsagittal CC.}}
\label{fig:IXIresults}
\end{figure}

\rev{Qualitatively, the unthresholded $g$ maps of the parameterization and TBSS showed age-related differences in the same direction in roughly the same regions, including localized areas with effects opposite to the global trend (e.g., higher FA and lower RD in older adults in the central CC) (Fig.{~\ref{fig:IXIresults}}a). The VBM-style analysis agreed with both methods in most of the CC, especially for AxD, but diverged in specific regions. In the central CC, for instance, it did not reproduce the opposite-direction effects in FA and RD. The parcellation followed the overall pattern but, given its regional resolution, did not show these localized differences.}

\rev{On the skeleton, the spatial correlation between $g$ maps was moderate to strong for all pairs of methods (Pearson’s $r$ = 0.59-0.82; Table{~\ref{tab:IXI_spatial_agreement}}) and slightly higher for VBM-style $\times$ TBSS than for parameterization $\times$ TBSS in all four DTI maps. The pairs differed mainly in agreement in scale and location: $C_b$ was around 0.95 for parameterization $\times$ TBSS in all four maps, compared with 0.41-0.69 for the other pairs, so Lin’s CCC was highest for parameterization $\times$ TBSS (0.62-0.74). Within the full CC mask, agreement between the parameterization and VBM-style was highest for AxD and lowest for FA and RD.}

Statistically significant results for FA and MD using TBSS were highly consistent with those reported by \citet{sheriff_comparison_2024} (Fig.~\ref{fig:IXIresults}b). All approaches identified significant age-related decreases in FA and increases in MD, RD, and AxD. For the parameterization, TBSS, and parcellation approaches, FA changes were concentrated in anterior regions, MD and RD changes extended across anterior and central regions, and AxD changes were concentrated in central regions. \rev{On the skeleton, the overlap between significant regions was low to moderate for all pairs of methods (Dice = 0.45-0.73; Table{~\ref{tab:IXI_spatial_agreement}}), with no pair consistently highest across DTI maps. Within the full CC mask, the overlap between the parameterization and VBM-style analysis was higher for MD, RD, and AxD and lower for FA. The parameterization also showed significant differences near the CC boundaries, outside the TBSS skeleton, as well as in localized regions of the posterior CC that were not significant with TBSS or the parcellation.}

% -------------------------------------------------
% -------------------------------------------------

% -------------------------------------------------
\begin{table}[ht]
\centering
\caption{\rev{Spatial similarity and agreement between methods for the comparison between younger and older adults in the IXI dataset. Dice coefficient was computed on the regions surviving FWE-corrected $p<0.05$, and Pearson’s $r$, $C_b$, and Lin’s CCC on the unthresholded $g$ maps. Higher values indicate greater similarity or agreement. Results were computed within the TBSS skeleton of the midsagittal CC (CC skeleton) or within the final CC mask from the parameterization (CC mask). Param: parameterization; VBM: VBM-style analysis. Values in bold indicate the highest value for the analysis within the skeleton.}}
\small
\begin{tabular}{llcccc}
\toprule
& & \multicolumn{3}{c}{CC skeleton} & \multicolumn{1}{c}{CC mask} \\
\cmidrule(lr){3-5} \cmidrule(lr){6-6}
 &  & VBM $\times$ TBSS & Param $\times$ TBSS
& Param $\times$ VBM & Param $\times$ VBM \\
\midrule
\multirow{4}{*}{FA}
& Dice coefficient & \textbf{0.646} & 0.597 & 0.464 & 0.534 \\
& Pearson's $r$ & \textbf{0.785} & 0.710 & 0.651 & 0.541 \\
& Bias correction factor $C_b$ & 0.592 & \textbf{0.955} & 0.671 & 0.610 \\
& Lin's CCC & 0.464 & \textbf{0.678} & 0.437 & 0.330 \\
\midrule
\multirow{4}{*}{MD}
& Dice coefficient & 0.532 & \textbf{0.725} & 0.650 & 0.855 \\
& Pearson's $r$ & \textbf{0.744} & 0.653 & 0.669 & 0.755 \\
& Bias correction factor $C_b$ & 0.441 & \textbf{0.945} & 0.460 & 0.624 \\
& Lin's CCC & 0.328 & \textbf{0.617} & 0.308 & 0.471 \\
\midrule
\multirow{4}{*}{RD}
& Dice coefficient & 0.558 & 0.619 & \textbf{0.681} & 0.779 \\
& Pearson's $r$ & \textbf{0.716} & 0.689 & 0.592 & 0.620 \\
& Bias correction factor $C_b$ & 0.411 & \textbf{0.953} & 0.459 & 0.553 \\
& Lin's CCC & 0.294 & \textbf{0.656} & 0.271 & 0.343 \\
\midrule
\multirow{4}{*}{AxD}
& Dice coefficient & 0.453 & 0.533 & \textbf{0.651} & 0.792 \\
& Pearson's $r$ & 0.788 & 0.776 & \textbf{0.824} & 0.847 \\
& Bias correction factor $C_b$ & 0.689 & \textbf{0.950} & 0.612 & 0.739 \\
& Lin's CCC & 0.543 & \textbf{0.737} & 0.504 & 0.625 \\
\bottomrule
\end{tabular}
\label{tab:IXI_spatial_agreement}
\end{table}
% -------------------------------------------------

In the VBM-style analysis, significant differences were observed across most of the CC, except for portions of the anterior and posterior CC (Fig.~\ref{fig:IXIresults}b). The spatial extent of these differences varied with the level of smoothing, with smaller clusters at lower smoothing levels and more widespread differences at higher smoothing levels (Supplementary Fig.~6). The VBM-style analysis identified differences extending beyond the CC parameterization mask, and both TBSS and VBM showed group differences outside the CC, including the fornix (Supplementary Figs.~6 and~7). The parcellation analysis using all voxels within each region showed higher effect sizes and more CC regions with significant group differences (Supplementary Fig.~8).

\rev{Regarding the sensitivity to input mask, group comparison results were largely consistent across the three TractSeg mask types (original, eroded, and dilated). Qualitatively, the unthresholded $g$ maps showed the same overall pattern across mask types (Supplementary Fig.~9). Quantitatively, Pearson’s $r$, $C_b$, and Lin’s CCC were at least 0.9 in all pairwise comparisons and DTI maps (Supplementary Table~5). The overlap between significant regions was high for MD, RD, and AxD (Dice above 0.89) and somewhat lower for FA (Dice = 0.73-0.79).}

\subsubsection{Alzheimer's Disease Patients \textit{vs}. Cognitively Normal Adults}

The final sample from the ADNI dataset comprised 30 AD participants (13 females, $72.2 \pm 8.9$ years) and 59 CN participants (28 females, $72.1 \pm 8.9$ years).

The average length of the centerline ($L$) was $77.75$ voxels ($97.19$ mm), while the average length of the transverse straight lines ($T$) was $7.34$ voxels ($9.18$ mm). This yielded a ratio $r = 11$. Nevertheless, the coverage of the initial sampling configuration ($(25\times 225)$ points) was $99.15\%$, whereas the coverage of the final sampling configuration ($(17\times 201)$ points) was $70.90\%$.

\begin{figure}[htb!]
  \centering
  \begin{tabular}{@{} m{0.8cm} @{} m{17cm} @{}}
    \small \textbf{(a)} & \includegraphics[width=17cm]{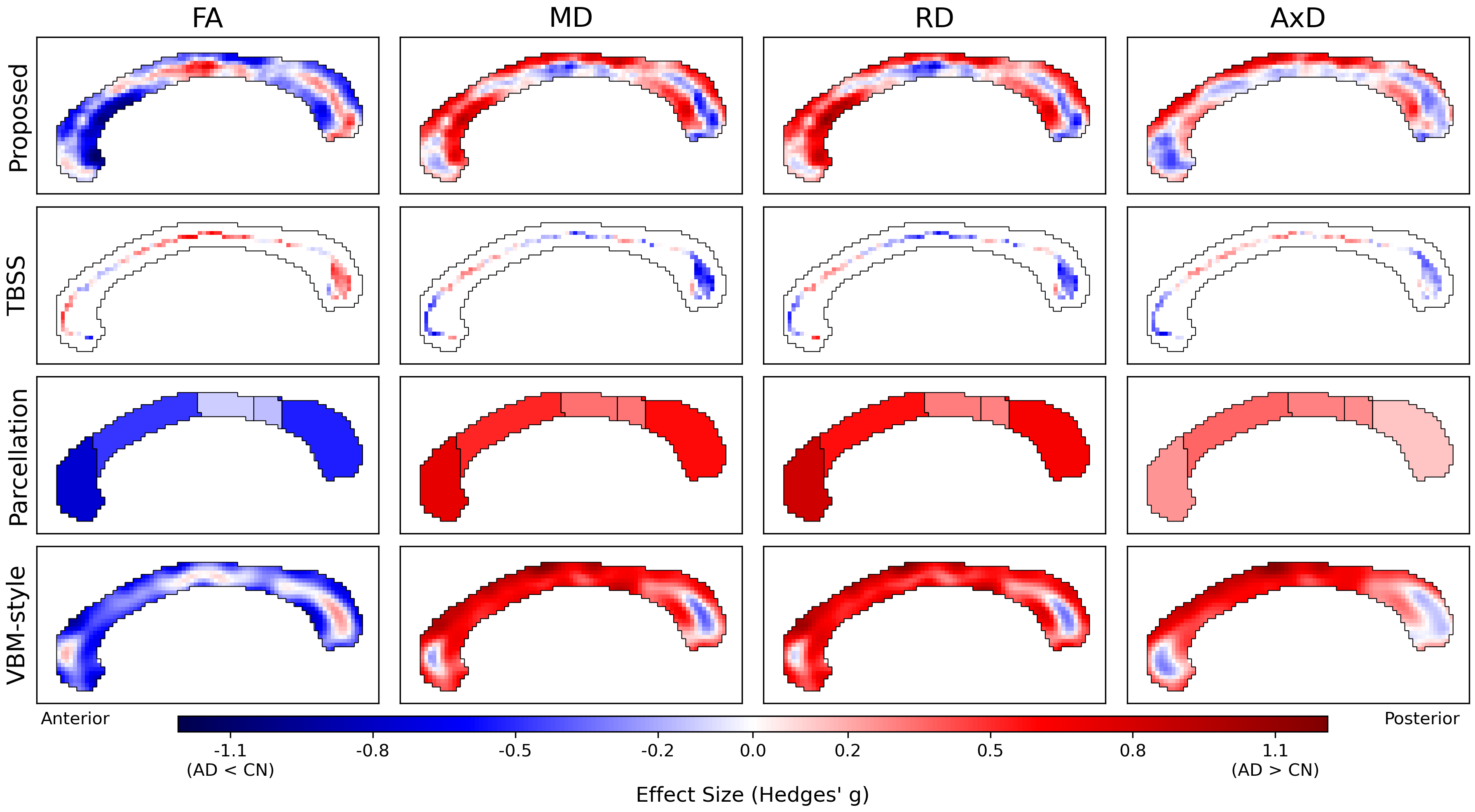} \\
    \small \textbf{(b)} & \includegraphics[width=17cm]{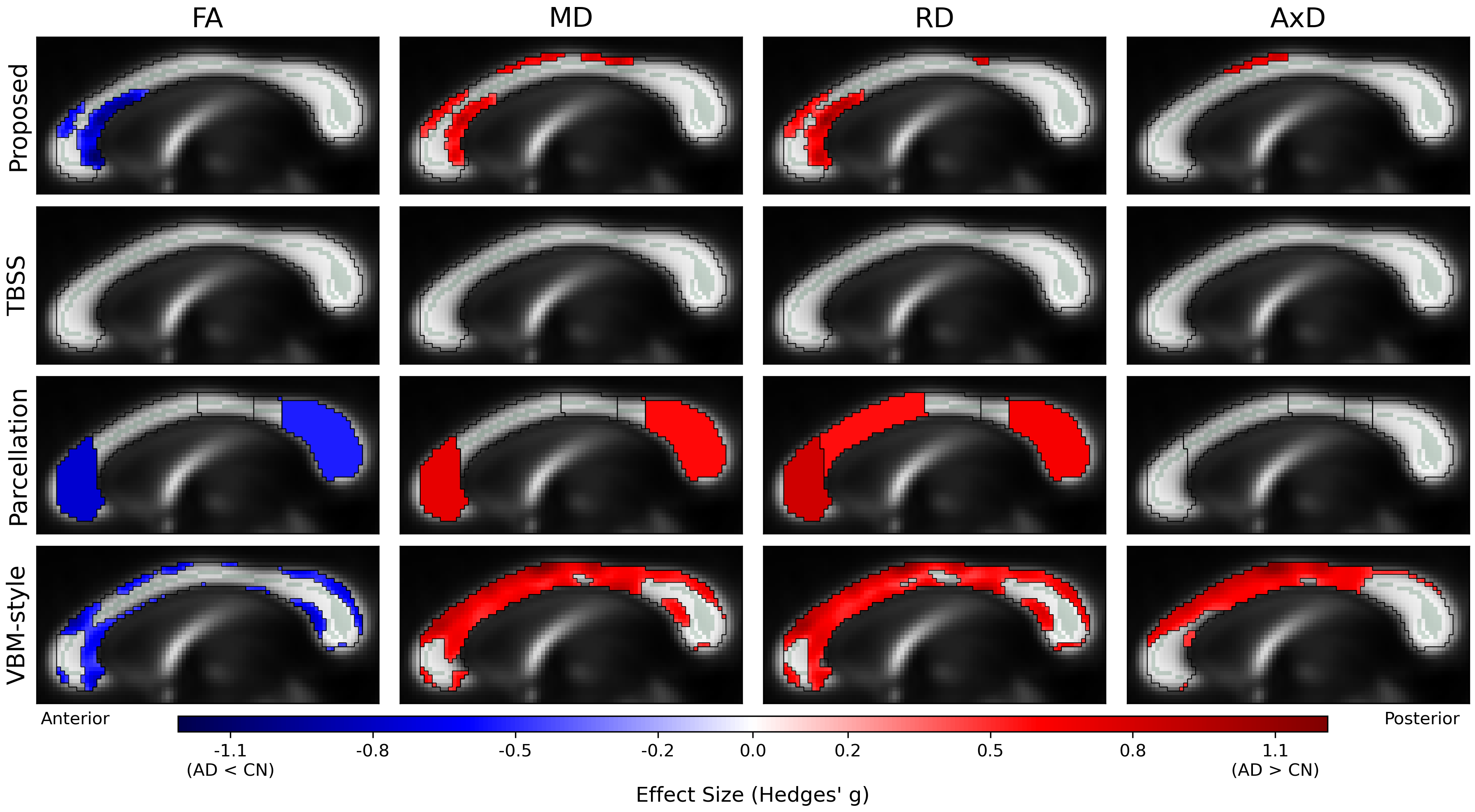}
  \end{tabular}
\caption{\rev{Group differences between Alzheimer’s disease (AD) and cognitively normal (CN) controls in the ADNI dataset, shown as Hedges’ $g$ maps. \textbf{(a)} Unthresholded maps. \textbf{(b)} Maps restricted to regions surviving FWE-corrected $p<0.05$ over mean FA image. Results are organized by method (rows) and DTI map (columns). TBSS and VBM-style results are shown within the CC parameterization mask (black contour). The grayish-green region denotes the TBSS skeleton of the midsagittal CC.}}
\label{fig:ADNIresults}
\end{figure}

\rev{The unthresholded $g$ maps of the parameterization and TBSS still showed a similar overall pattern, but with more disagreements than in the IXI analysis (Fig.{~\ref{fig:ADNIresults}}a). Visually, the anterior portion of the skeleton showed clear FA increases and MD and RD decreases in AD relative to CN with TBSS, whereas these trends were less clear in the parameterization results, which presented lower effect sizes and some effects in the opposite direction. The VBM-style analysis agreed partially with the other methods, diverging in specific regions, such as the central CC, where it did not show the MD and RD decreases seen with the parameterization and TBSS. The parcellation followed the overall pattern but did not show the localized differences.}

\rev{On the skeleton, the spatial correlation between the parameterization and TBSS was moderate (Pearson’s $r$ = 0.55-0.59; Table{~\ref{tab:ADNI_spatial_agreement}}), lower than in IXI and than that between the VBM-style analysis and TBSS (0.64-0.80). Agreement in scale and location again favored parameterization $\times$ TBSS, so Lin’s CCC was higher for parameterization $\times$ TBSS than for VBM-style $\times$ TBSS in all four maps. For FA, $C_b$ and CCC were highest for parameterization $\times$ VBM-style. Within the full CC mask, agreement between the parameterization and VBM-style was low to moderate.}

Using TBSS, no statistically significant differences were found in the midsagittal CC skeleton (Fig.~\ref{fig:ADNIresults}b).
In the most anterior CC region, the parameterization, parcellation, and VBM-style approaches yielded concordant findings, showing significant FA decreases and MD and RD increases in AD compared with CN individuals. The parameterization further detected localized MD, RD, and AxD alterations near the superior boundary of the central CC that were not captured by the parcellation. Conversely, the parcellation identified significant FA, MD, and RD differences in the most posterior CC region, as well as RD alterations in the second most anterior region, which were not detected by the parameterization but were generally consistent with the VBM-style findings.
\rev{As TBSS had no significant voxels, the overlap between significant regions could only be assessed between the parameterization and the VBM-style analysis. It was low on the skeleton and for AxD within the CC mask, and low to moderate for the other DTI maps within the CC mask (Table{~\ref{tab:ADNI_spatial_agreement}}).}

% -------------------------------------------------
% -------------------------------------------------

In the VBM-style analysis, significant differences in MD, RD, and AxD were observed across most of the central CC, diverging from the parameterization and parcellation results. The spatial extent of these differences varied with the level of smoothing and extended beyond the CC parameterization mask, including the fornix (Supplementary Fig.~10). The parcellation approach using all voxels identified alterations in additional CC regions (Supplementary Fig.~11).

% -------------------------------------------------
\begin{table}[tb]
\centering
\caption{\rev{Spatial similarity and agreement between methods for the comparison between AD and CN individuals in the ADNI dataset. Dice coefficient was computed on the regions surviving FWE-corrected $p<0.05$, and Pearson’s $r$, $C_b$, and Lin’s CCC on the unthresholded $g$ maps. Higher values indicate greater similarity or agreement. Results were computed within the TBSS skeleton of the midsagittal CC (CC skeleton) or within the final CC mask from the parameterization (CC mask). Param: parameterization; VBM: VBM-style analysis. Values in bold indicate the highest value for the analysis within the skeleton. Dashes indicate that Dice could not be computed because at least one of the methods had no significant voxels within the skeleton.}}
\small
\begin{tabular}{llcccc}
\toprule
& & \multicolumn{3}{c}{CC skeleton} & \multicolumn{1}{c}{CC mask} \\
\cmidrule(lr){3-5} \cmidrule(lr){6-6}
 &  & VBM $\times$ TBSS & Param $\times$ TBSS
& Param $\times$ VBM & Param $\times$ VBM \\
\midrule
\multirow{4}{*}{FA}
& Dice coefficient & -- & -- & \textbf{0.118} & 0.424 \\
& Pearson's $r$ & \textbf{0.750} & 0.575 & 0.563 & 0.566 \\
& Bias correction factor $C_b$ & 0.534 & 0.794 & \textbf{0.871} & 0.887 \\
& Lin's CCC & 0.400 & 0.456 & \textbf{0.491} & 0.502 \\
\midrule
\multirow{4}{*}{MD}
& Dice coefficient & -- & -- & \textbf{0.085} & 0.473 \\
& Pearson's $r$ & \textbf{0.736} & 0.587 & 0.578 & 0.512 \\
& Bias correction factor $C_b$ & 0.454 & \textbf{0.898} & 0.598 & 0.718 \\
& Lin's CCC & 0.334 & \textbf{0.527} & 0.345 & 0.368 \\
\midrule
\multirow{4}{*}{RD}
& Dice coefficient & -- & -- & \textbf{0.267} & 0.415 \\
& Pearson's $r$ & \textbf{0.635} & 0.577 & 0.470 & 0.456 \\
& Bias correction factor $C_b$ & 0.414 & \textbf{0.840} & 0.635 & 0.704 \\
& Lin's CCC & 0.263 & \textbf{0.484} & 0.298 & 0.321 \\
\midrule
\multirow{4}{*}{AxD}
& Dice coefficient & -- & -- & -- & 0.153 \\
& Pearson's $r$ & \textbf{0.802} & 0.545 & 0.608 & 0.653 \\
& Bias correction factor $C_b$ & 0.536 & \textbf{0.988} & 0.478 & 0.705 \\
& Lin's CCC & 0.429 & \textbf{0.539} & 0.291 & 0.460 \\
\bottomrule
\end{tabular}
\label{tab:ADNI_spatial_agreement}
\end{table}
% -------------------------------------------------

\rev{Group comparison results were again largely consistent across the three TractSeg mask types. Qualitatively, the unthresholded $g$ maps showed the same overall pattern across mask types (Supplementary Fig.~12). Quantitatively, Pearson’s $r$, $C_b$, and Lin’s CCC were above 0.93 in all pairwise comparisons and DTI maps (Supplementary Table~6). The overlap between significant regions was high for FA, MD, and RD (Dice above 0.85) and could not be computed for AxD because there were no significant voxels within the intersection of the templates.}

% ===========================================================
\section{Discussion}
\label{sec:discussion}
% ===========================================================

Here, we propose a new analysis method using parameterization to resample the midsagittal section of the CC using anatomical references, producing a standardized grid with a fixed number of sampling points. By establishing approximate correspondence between these points across individuals, diffusion measurements can be compared directly despite natural variations in CC shape and size.

The motivation for developing such a framework stems from the limitations of existing approaches commonly used to study the CC in dMRI. VBM-style analyses allow detailed investigations of the whole brain, but rely heavily on accurate inter-subject registration and parameter choices, which can introduce distortions and biases. Although TBSS was designed to mitigate these issues by projecting the data onto a WM skeleton, it restricts the analysis to the central portion of WM tracts and may overlook alterations near tract boundaries. In contrast, CC parcellation approaches are widely used for region-specific studies of the structure, but they rely on averaging diffusion metrics within relatively large regions and lack a standardized framework, which limits spatial detail and complicates comparisons across studies. The proposed parameterization provides a simple and computationally efficient alternative for the detailed characterization of the midsagittal CC with less sensitivity to registration. To facilitate its adoption, the code for the final configuration of the parameterization method is openly available at \url{https://github.com/MICLab-Unicamp/Midsagittal-CC-Parameterization} (under preparation).

\subsection{Configuration and Repeatability Experiments}

The repeatability analyses performed on the HCP-TR dataset demonstrated the consistency of the proposed method. From eleven transverse points onward, variation across sampling configurations was minimal (Fig.~\ref{fig:npoints}), indicating that the method would remain stable even with changes in parameter selection. Average repeatability metrics were consistently below the commonly accepted threshold of $10\%$~\citep{marenco_regional_2006}, further supporting the robustness of the approach. Although configurations with fewer sampling points yielded similar repeatability, we selected a denser configuration with 25 transverse points to ensure comprehensive CC coverage, particularly considering potential variability in CC size across datasets. Consequently, even with the variation in CC size and proportions observed in the ADNI dataset, the resulting coverage remained comparable to that obtained for the HCP-TR and IXI datasets.

When restricting the analysis to the parameterization centerline, repeatability metrics were comparable to those obtained with TBSS (Tables~\ref{tab:error} and~\ref{tab:cov}), indicating that the proposed framework preserves the integrity of the diffusion information in a manner consistent with established approaches. TBSS showed better repeatability for FA, consistent with its FA-based skeleton construction. However, this advantage did not extend to MD, RD, and AxD, for which the parameterization centerline performed similarly or better. It is important to note that the parameterization centerline and the midsagittal CC skeleton are not equivalent. They differ both in their position in template space (Fig.~\ref{fig:rows_cols_addition}a) and in the way their associated values are computed. Nevertheless, this analysis confirmed that the parameterization does not introduce artifacts and yields reliable results in regions where robustness is typically higher.

Although restricting the analysis to the centerline would maximize repeatability, it would substantially limit CC coverage (Fig.~\ref{fig:rows_cols_addition}c). Therefore, we progressively added pairs of longitudinal and transverse curves to evaluate the trade-off between coverage and repeatability. In the longitudinal direction, eight curves provided a suitable balance, achieving approximately $70\%$ coverage while remaining below half of the total repeatability loss observed across configurations. In contrast, adding transverse curves contributed little to overall coverage because the points become progressively closer to one another. Consequently, we limited the number of transverse curves and stopped adding them once the metrics began to increase. This strategy allowed us to select a configuration that preserves most of the CC while avoiding substantial losses in measurement stability.

The repeatability achieved by the final configurations was lower than that obtained with the centerline alone or with TBSS (Tables~\ref{tab:error} and~\ref{tab:cov}). Compared with the centerline results, RD was the most affected map, showing an increase of approximately $2\%$ in the average metrics, whereas FA exhibited the largest increase in variability, as reflected by the standard deviation. Nevertheless, the repeatability metrics remained below $10\%$ for all diffusion maps except RD. In fact, RD consistently showed the lowest repeatability and the highest standard deviation across all evaluated scenarios, with average values exceeding $10\%$ even for TBSS and the centerline. This behavior may be attributed to its dependence on the two minor eigenvalues of the diffusion tensor, which are more sensitive to noise than the principal eigenvalue, especially in highly anisotropic regions~\citep{pierpaoli_toward_1996}.

Regarding the transverse curves, we evaluated two strategies to determine whether a simple definition (straight lines) would be sufficient to generate a reliable parameterization or whether a more complex grid, better adapted to the curved CC shape (Laplace streamlines), would provide more robust results. The straight-line strategy produced a more homogeneous sampling distribution by placing equally spaced points along the CC boundaries. In contrast, the Laplace streamlines were closer together in the central portion of the CC and more separated in the anterior and posterior regions, particularly near the superior boundary. This resulted in greater variation in sampling density (Fig.~\ref{fig:selec_pts_distribution}), requiring more streamlines to avoid gaps in some regions. The Laplace method was also substantially more computationally demanding. Such a cost could be justified if it yielded higher repeatability, but the results were nearly identical for both strategies (Tables~\ref{tab:error} and~\ref{tab:cov}), with worse average metrics for the Laplace method, possibly due to its slightly higher coverage. Considering these factors, the straight-line strategy was selected as the final parameterization configuration.

\rev{Because the parameterization relies on CC segmentation, we analyzed the method’s sensitivity to the input mask by eroding and dilating the original TractSeg masks, as well as by replacing them with masks obtained with MRICloud. As expected, larger masks tended to result in worse repeatability metrics (Table{~\ref{tab:input_summary}}; Supplementary Tables~1 and~2), likely because they include more peripheral voxels, which are more prone to partial volume effects. When the analysis was restricted to the voxels present in the templates of all mask types, the range of the metrics across mask types fell by roughly three-quarters (Table{~\ref{tab:input_summary}}; Supplementary Tables~3 and~4), indicating that most of the difference in repeatability arises from the additional peripheral voxels. In these common voxels, the remaining differences reflect how the mask alters the parameterization itself and were small: the relative differences of the dilated and eroded masks compared with the original fell from $+21\%$ and $-10\%$ to $+3\%$ and $-3\%$, respectively. The MRICloud masks presented the largest difference from the original in the common voxels (relative increases of $4$-$5\%$) and the worst repeatability among mask types. This may reflect the difference in how the masks are obtained: MRICloud segments the CC on T1w images using a multi-atlas approach and required resampling of the masks to diffusion space, whereas TractSeg operates directly on diffusion images with a deep-learning approach, and the two may also define the CC boundaries differently. Together, these results indicate that, in the central CC, the repeatability of the parameterization is robust to the input mask variations tested here (boundary shifts of one voxel and an alternative segmentation), whereas peripheral voxels are more sensitive, which is consistent with the rationale for excluding the outermost parameterization points.}

\rev{The MRICloud masks required manual correction in four individuals of the HCP-TR dataset, suggesting that its segmentation was less reliable than TractSeg’s in this dataset. When comparing the uncorrected and corrected masks of these individuals, correction improved the repeatability measures for FA and AxD (relative decreases of $6.5\%$ to $8.5\%$ in the mean values, with smaller standard deviations), whereas MD and RD were essentially unchanged (Table{~\ref{tab:corrected_uncorrected}}). Overall, results were similar between the two versions of the masks, suggesting limited sensitivity of the parameterization to minor localized differences in the input masks, although this comparison comprised only four individuals. It is important to highlight that only minor corrections were necessary (Supplementary Fig.~5), and more extensive segmentation inaccuracies could have a larger impact. Therefore, as with any application that requires segmentation, quality control is necessary to ensure the correctness of the masks and underlying analysis, and we recommend assessing the CC masks and the parameterization-derived boundaries together to make sure everything is working as expected. Given that the MRICloud masks differed more from the original masks in the common voxels and required manual correction more often, we suggest using the parameterization with TractSeg or another segmentation method that operates directly on diffusion data, which also avoids resampling between image spaces.}

\rev{Finally, it is important to note that repeatability quantifies measurement stability, that is, the consistency of the sampled values across repeated scans of the same individuals, and is therefore a necessary condition for detecting subtle differences. It does not establish biological validity: stable values are not necessarily accurate or specific to the tissue properties of interest, and the analysis does not address whether the same locations correspond across individuals (see Section~}\ref{sec:limitations}).

\subsection{Example Applications}

\rev{The two DTI-based group analyses used datasets that differ from the one in which the method was configured (HCP-TR: $1.25$ mm isotropic, 90 directions): IXI ($1.75 \times 1.75 \times 2.0$ mm$^3$, 15 directions) and ADNI ($2$ mm isotropic, 48 directions). After resampling to $1.25$ mm, the same pipeline and parameters could be applied to both, indicating that the framework is applicable to acquisitions with different protocols and resolutions.} The results highlighted both the consistency of the proposed method relative to established approaches and its ability to provide more spatially detailed information about CC alterations. Across analyses, the parameterization produced results that were broadly consistent with TBSS and parcellation-based approaches while offering a more continuous representation of diffusion changes across the midsagittal CC. These comparisons illustrate how the proposed framework can complement existing methods by combining simplicity with increased spatial resolution.

In the age-group comparison, we replicated a study originally conducted using TBSS, which provided well-defined expected findings. Although a different dataset was used, our TBSS results were highly consistent with those reported by \citet{sheriff_comparison_2024}, with FA decreases and MD increases appearing in similar CC regions. This agreement increases confidence in the validity of our analysis and provides a reference for comparing the approaches, \rev{although it is not a ground truth for the spatial distribution of age-related differences. The AD-CN comparison, in contrast, was not intended to replicate specific findings, and TBSS found no significant differences in the midsagittal CC skeleton.}

\rev{In both datasets, the unthresholded effect-size maps of the parameterization and TBSS showed a similar overall pattern, with more disagreement in ADNI, where effects are subtler. The spatial correlation with TBSS was higher for the VBM-style analysis than for the parameterization in all DTI maps, but agreement in scale and location was markedly higher for the parameterization, so Lin’s CCC favored it in every map (Tables{~\ref{tab:IXI_spatial_agreement}} and{~\ref{tab:ADNI_spatial_agreement}}). The VBM-style and TBSS maps thus share a broad spatial pattern but differ in the magnitude of effect sizes, whereas the parameterization reproduced the magnitude of TBSS to a greater extent while retaining a moderate to strong spatial pattern. As TBSS and the VBM-style analysis used the same steps to register the images to a common space, a high correlation between them is expected and is not independent evidence of agreement.
The low $C_b$ is likely a result of the smoothing applied in the VBM-style analysis, which can attenuate peak effect sizes and compress the variance, consistent with the dependence of its results on the smoothing level (Supplementary Figs.~6 and~10)
Likewise, most localized effects opposite to the global trend seen with the parameterization and TBSS were absent from the VBM-style maps, as expected if smoothing blurs localized features. The parameterization does not apply smoothing or require registration of the images to a common space, while providing the spatial continuity across the midsagittal CC that the skeleton and the parcellation lack.}

\rev{Agreement between significant regions was lower than between the unthresholded maps, and no pair of methods was consistently highest, although in IXI the parameterization tended to overlap more with TBSS than the VBM-style analysis did, except for FA (Tables{~\ref{tab:IXI_spatial_agreement}} and{~\ref{tab:ADNI_spatial_agreement}}). The methods thus agree more on the overall topography of the differences than on which voxels reach significance, which also depends on search volume, smoothing, and sensitivity.}
In IXI, the parameterization showed additional differences near the CC boundaries, \rev{consistent with the restriction of TBSS to a narrow skeleton, which in the midsagittal CC leaves much of the structure unsampled.} In ADNI, despite the low Dice coefficients, the anterior differences detected by the parameterization and VBM-style analyses were located closer to the CC boundaries, which may explain why TBSS did not detect them.

The parameterization may also help interpret discrepancies between TBSS and the parcellation. In IXI, for example, TBSS detected little evidence of MD differences in the most anterior CC region, whereas the parcellation indicated significant changes. By averaging values within predefined regions, the parcellation cannot distinguish whether differences are localized or widespread. In the parameterization, these differences were concentrated near the region boundaries, which is consistent with differences widespread enough to influence the regional mean but away from the high-FA core sampled by TBSS. Similarly, the parameterization identified localized alterations in the posterior CC that were both away from the local FA maxima and not extensive enough to influence either TBSS or the parcellation. Conversely, in ADNI, the parcellation indicated additional posterior differences, consistent with the VBM-style findings but not detected by the parameterization, which may reflect the higher sensitivity of regional averages to diffuse, low-amplitude differences, whereas voxelwise inference requires larger effects to survive correction.

The VBM-style analysis is perhaps the most directly comparable to the parameterization, as both are voxel-based and evaluate the entire CC. In both datasets, however, it indicated more widespread differences: in IXI, particularly for FA, \rev{for which its overlap with the parameterization within the CC mask was lowest}, and in ADNI, with significant AxD differences across nearly the entire central CC that the other methods did not show. In several cases, these differences spanned the entire CC thickness, necessarily including the local FA maxima, yet TBSS found no significant differences in those regions. Importantly, the spatial extent of the VBM-style results was highly sensitive to the degree of smoothing applied (Supplementary Figs.~6 and~10). Registration quality may also influence these results, as imperfect alignment can introduce systematic biases (e.g., related to brain size or morphology)~\citep{bookstein_voxel-based_2001,ashburner_why_2001}. This sensitivity, combined with the agreement in the significant regions observed among the parameterization, TBSS, and parcellation approaches, raises concerns about the reliability of the VBM-style findings in this context.

Even though the AD-CN analysis was not intended to replicate specific prior findings, numerous DTI studies have reported CC alterations associated with AD. A meta-analysis by \citet{sexton_meta-analysis_2011}, which focused on region-of-interest-based studies, reported medium effect sizes in both the genu (anterior region) and splenium (posterior region) of the CC for FA and MD when comparing AD patients with controls, which is in line with our parcellation results. Similarly, \citet{kumar_topology_2022} performed a region-based analysis of CC segments using DTI maps and additional diffusion metrics, reporting significant differences between AD and control groups in the mid-anterior and mid-posterior CC for FA, MD, and RD, as well as alterations in the anterior and mid-anterior regions for AxD. Although their parcellation scheme differs from ours and prevents a direct comparison, the strongest effects were observed in the mid-anterior CC, consistent with the most prominent alterations detected by the proposed parameterization.

\rev{The group-level results obtained with the parameterization were also robust to the input mask. In both datasets, the unthresholded effect-size maps were nearly interchangeable across the original, eroded, and dilated TractSeg masks, and the significant regions overlapped largely (Dice above 0.85 for all maps with significant voxels, except FA in IXI, with 0.73-0.79). Thus, a one-voxel shift in the mask boundary, which noticeably affected the repeatability of peripheral values, had little effect on the central group differences. The lower overlap for FA in IXI was not investigated, but significance maps are expected to differ more than effect-size maps, as small changes in voxels near the significance threshold alter the extent of the significant regions.}

It is important to note that the evaluated methods rely on different assumptions, and therefore their results are not strictly equivalent. TBSS, for example, was designed for whole-brain WM analysis and involves image registration, skeletonization, and projection of FA values onto the WM skeleton. Consequently, a significant voxel in the TBSS skeleton does not necessarily correspond to the exact anatomical location sampled by the parameterization. Nevertheless, the overall agreement among the parameterization, TBSS, and parcellation approaches indicates that the proposed framework captures meaningful diffusion patterns while providing a more spatially detailed description of CC alterations.

Given the topographic organization and functional specialization of the midsagittal CC, regional changes in diffusion metrics may reflect alterations in specific interhemispheric cortical connections. In this context, the proposed parameterization may offer new perspectives for studying CC variability, both in normal aging and in neurological conditions that affect the structure.

% \subsection{Study Limitations}
\subsection{Limitations and Future Work}
\label{sec:limitations}

Despite the promising results, some limitations should be considered. The proposed parameterization depends on accurate segmentation of the CC, as the method uses the geometry and boundaries of the structure to define the sampling grid. Consequently, segmentation errors may propagate to the subsequent analysis and affect the spatial localization of the measurements.
\rev{We assessed this dependency with one-voxel boundary shifts, an alternative segmentation method, and manually corrected versus uncorrected masks. In the central CC, these perturbations had a small effect on repeatability and on the group-level results, whereas peripheral voxels were more sensitive. More extensive segmentation errors could have larger effects, so quality control of the masks and of the parameterization-derived boundaries remains necessary, as discussed above.}

Another limitation relates to the assumption of correspondence between sampled points across individuals. The parameterization establishes geometric correspondence, that is, a consistent relative position within the midsagittal CC, by resampling the structure according to its shape. This does not guarantee biological homology: points with the same coordinates may not represent the same fiber populations in different individuals, particularly in the presence of anatomical variability, aging, atrophy, or pathology-related deformation -- a caveat common to most approaches that attempt to standardize anatomical structures.
\rev{The mask sensitivity analysis indicates that the sampled values in the central CC do not depend strongly on the exact position of the boundary, but it does not validate homology, and the agreement of the group-level regional patterns with TBSS and the parcellation is only indirect evidence. Direct validation, for example against histology, is not feasible with the available data. Regional findings should therefore be interpreted in terms of geometric position, and future work should evaluate whether the parameterized locations are consistent in their connectivity profiles across individuals, particularly in populations with pronounced structural alterations.}

Additionally, the method focuses exclusively on the midsagittal section of the CC and therefore does not capture alterations in other portions of the structure or in different WM tracts. For this reason, the parameterization should be viewed as complementary to whole-brain approaches rather than a replacement. \rev{The same idea of standardized sampling could in principle be adapted to other tracts or structures whose geometry allows a comparable parameterization, but this requires defining a suitable sampling grid for each structure and was not evaluated here.}

Finally, all experiments were conducted using a consistent preprocessing pipeline, and the robustness of the method to alternative preprocessing strategies remains to be investigated. \rev{The framework was applied to three datasets that differ in acquisition protocol and spatial resolution, and all data were resampled to $1.25$ mm isotropic before computing the DTI maps as part of the pipeline. The parameters were defined in HCP (native $1.25$ mm resolution) and applied unchanged to IXI and ADNI, which have coarser resolution and fewer directions. The group analyses of these datasets yielded results broadly in line with prior findings, but repeatability was assessed only in HCP. Therefore, the robustness of the method to scanner, protocol, and resolution was not evaluated systematically, and multi-site validation, for example with traveling subjects, remains to be done.}

\rev{Although we evaluated the framework only with DTI scalar maps, the parameterization itself does not depend on the metric: any scalar map defined voxelwise in the midsagittal plane and aligned with the CC mask can be sampled at the same points. Examples are the neurite density and orientation dispersion indices from NODDI~}\citep{zhang_noddi_2012}\rev{, axon diameter indices from AxCaliber~}\citep{assaf_axcaliber_2008}\rev{ or ActiveAx~}\citep{alexander_orientationally_2010}\rev{, and even maps from other MRI contrasts. Given the regional variations in axon caliber and fiber composition across the CC, the spatial detail provided by the parameterization could be particularly valuable for these metrics.
Fixel-based measures are not voxelwise scalars and require dedicated statistical frameworks~}\citep{raffelt_connectivity-based_2015}\rev{, because neighboring voxels can contain different numbers of fixels, which may belong to different tracts. Applying the parameterization to them would require either aggregating them per voxel, which may be reasonable in the midsagittal CC, where a single fiber population dominates, or adapting the mapping and inference.
Still, because advanced acquisitions often have lower resolution or signal-to-noise ratio and some estimates (e.g., axon diameter) are noisy and require strong gradients, these extensions remain to be demonstrated.}

% ===========================================================
\section{Conclusions}
\label{sec:conclusion}
% ===========================================================

This paper introduced a parameterization framework for the analysis of the midsagittal CC in dMRI that \rev{samples diffusion measures at geometrically corresponding locations across individuals} without explicit inter-subject registration. In aging and Alzheimer’s disease, its results were broadly in line with TBSS, parcellation-based analyses, and the existing literature, while offering a more continuous and spatially detailed view of where differences are located along the CC. \rev{These findings support its use as a structure-specific complement to whole-brain and region-based methods. Its application to other diffusion metrics and MRI contrasts, and its systematic evaluation in multi-site data, are directions for future work.} By enabling spatially detailed mapping of diffusion measures along the CC, the framework may help characterize microstructural differences in aging and in neurological conditions that affect the CC.

\section*{Ethics}

The work presented in this manuscript was conducted using open-access data from the Human Connectome Project, IXI, and Alzheimer’s Disease Neuroimaging Initiative.

\section*{Data and Code Availability}

The final configuration of the parameterization method is publicly available on GitHub (\url{https://github.com/MICLab-Unicamp/Midsagittal-CC-Parameterization}; repository under preparation and will be finalized upon acceptance). The brain imaging datasets from the Human Connectome Project, IXI, and the Alzheimer’s Disease Neuroimaging Initiative are publicly available from their respective repositories, subject to each dataset's data use agreement.

\section*{Author Contributions}

% Author Contributions text (mandatory).

Caio Santana: Conceptualization, Methodology, Software, Validation, Formal analysis, Investigation, Data Curation, Writing - Original Draft, Visualization, Project administration. Catherine Lebel: Validation, Writing - Review \& Editing. Richard Frayne: Conceptualization, Methodology, Resources, Writing - Review \& Editing. Letícia Rittner: Conceptualization, Methodology, Resources, Writing - Review \& Editing, Supervision, Project administration, Funding acquisition.

\section*{Funding}

This study was partially supported by the São Paulo Research Foundation (FAPESP) grants \#2013/07559-3 and \#2020/09838-0. L. Rittner thanks the National Council of Scientific and Technological Development (CNPq grants \#317133/2023-3 and \#444725/2024-5). C. Santana thanks CNPq (grant \#141079/2021-5) and the Coordenação de Aperfeiçoamento de Pessoal de Nível Superior - Brasil (CAPES) - Finance Code 001 for his scholarships.

\section*{Declaration of Competing Interests}

% Declaration of Competing Interests text (mandatory).

The authors declare no competing interests.

\section*{Acknowledgements}

% Acknowledgements text (optional).

Data were provided in part by the Human Connectome Project, WU-Minn Consortium (Principal Investigators: David Van Essen and Kamil Ugurbil; 1U54MH091657) funded by the 16 NIH Institutes and Centers that support the NIH Blueprint for Neuroscience Research; and by the McDonnell Center for Systems Neuroscience at Washington University.

Part of the data used in preparing this article were obtained from the ADNI database (adni.loni.usc.edu). Data collection and sharing for this project was funded by the Alzheimer's Disease Neuroimaging Initiative (ADNI) (National Institutes of Health Grant U01 AG024904) and DOD ADNI (Department of Defense award number W81XWH-12-2-0012). ADNI is funded by the National Institute on Aging, the National Institute of Biomedical Imaging and Bioengineering, and through generous contributions from the following: AbbVie, Alzheimer’s Association; Alzheimer’s Drug Discovery Foundation; Araclon Biotech; BioClinica, Inc.; Biogen; Bristol-Myers Squibb Company; CereSpir, Inc.; Cogstate; Eisai Inc.; Elan Pharmaceuticals, Inc.; Eli Lilly and Company; EuroImmun; F. Hoffmann-La Roche Ltd and its affiliated company Genentech, Inc.; Fujirebio; GE Healthcare; IXICO Ltd.; Janssen Alzheimer Immunotherapy Research \& Development, LLC.; Johnson \& Johnson Pharmaceutical Research \& Development LLC.; Lumosity; Lundbeck; Merck \& Co., Inc.; Meso Scale Diagnostics, LLC.; NeuroRx Research; Neurotrack Technologies; Novartis Pharmaceuticals Corporation; Pfizer Inc.; Piramal Imaging; Servier; Takeda Pharmaceutical Company; and Transition Therapeutics. The Canadian Institutes of Health Research is providing funds to support ADNI clinical sites in Canada. Private sector contributions are facilitated by the Foundation for the National Institutes of Health (www.fnih.org). The grantee organization is the Northern California Institute for Research and Education, and the study is coordinated by the Alzheimer’s Therapeutic Research Institute at the University of Southern California. ADNI data are disseminated by the Laboratory for Neuro Imaging at the University of Southern California.

\section*{Declaration of the use of AI}

With respect to the research, AI was used for the automated segmentation of the corpus callosum, using the deep-learning-based tool TractSeg~\citep{wasserthal_tractseg_2018}, as described in Section~\ref{sec:datasets}. No other AI tools were used in planning or carrying out the research.

With respect to writing, AI was used to check spelling and grammar and to reword our draft text. The exact AI tool was Claude Opus 5.5 (Anthropic). AI was not used to generate figures.

\clearpage
\printbibliography

\clearpage
\includepdf[pages=-]{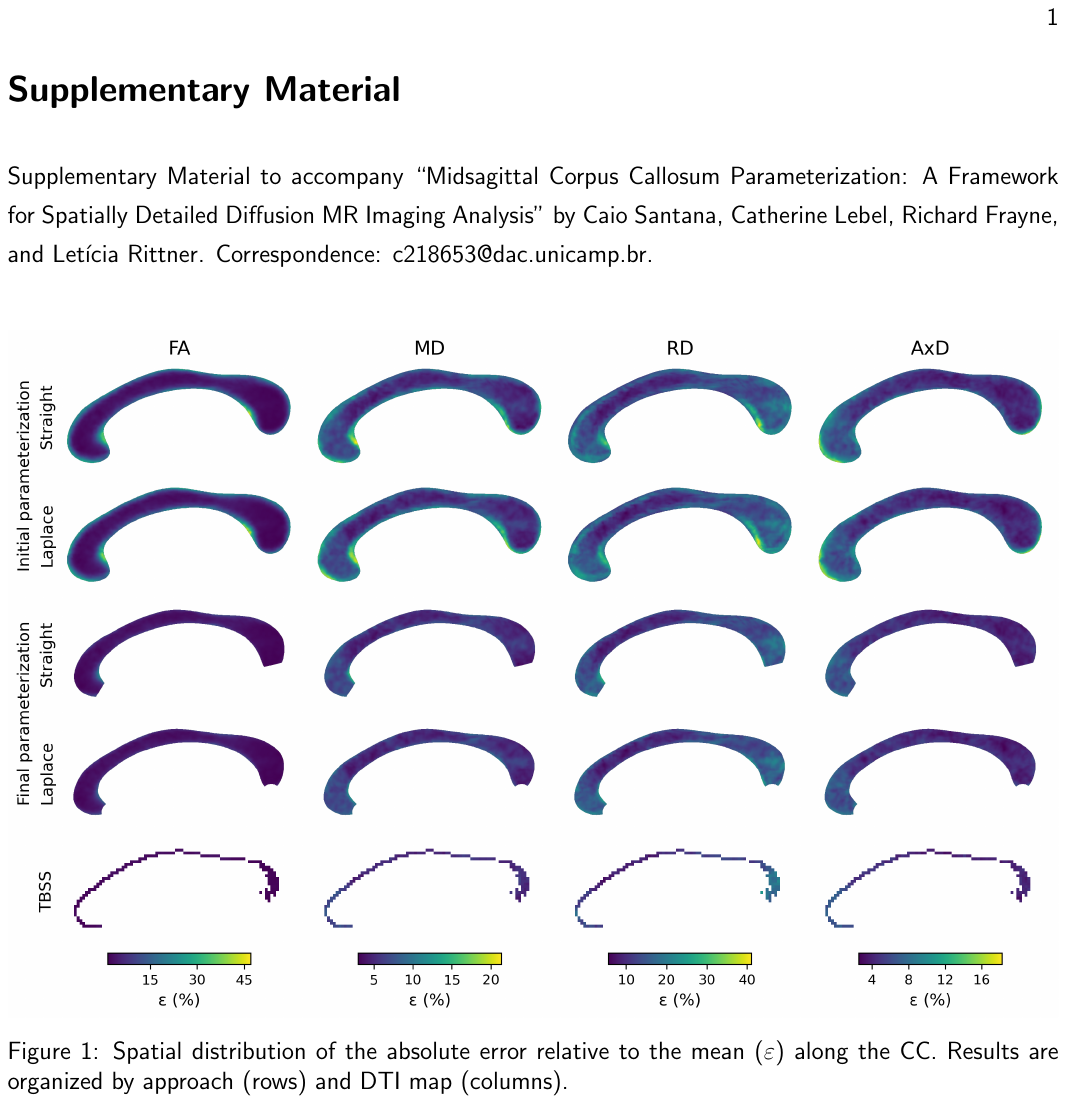}

\end{document}